\documentclass[fleqn,usenatbib]{mnras}
\usepackage{newtxtext,newtxmath}
\DeclareRobustCommand{\VAN}[3]{#2}
\let\VANthebibliography\thebibliography
\def\thebibliography{\DeclareRobustCommand{\VAN}[3]{##3}\VANthebibliography}
\usepackage{cite}
\usepackage{hyperref}
\usepackage{adjustbox}
\usepackage{blindtext}
\usepackage{multirow}
\usepackage{multicol}
\usepackage{cuted}
\usepackage{comment}
\usepackage{graphicx}	
\usepackage{amsmath}	

\title[FSRQ 3C\,345 as Very High Energy Blazar]{Can FSRQ 3C\,345 be a Very High Energy Blazar Candidate?}

\author[Dar Athar et al.]{
Athar A. Dar,$^{1,2}$\thanks{E-mail: ather.dar6@gmail}
Sunder Sahayanathan,$^{3,4}$\thanks{E-mail: sunder@barc.gov.in}
Zahir Shah,$^{2}$\thanks{E-mail: shahzahir4@gmail.com}
and Naseer Iqbal$^{1}$
\\
$^{1}$Department
of Physics, University of Kashmir, Srinagar 190006, India.\\
$^{2}$Department of Physics, Central University of Kashmir, Ganderbal 191201, India.\\
$^{3}$Astrophysical Sciences Division, Bhabha Atomic Research Center, Mumbai 400085, India.\\
$^{4}$Homi Bhabha National Institute, Mumbai 400094, India.}

\date{Accepted XXX. Received YYY; in original form ZZZ}

\pubyear{2023}

\begin{document}
\label{firstpage}
\pagerange{\pageref{firstpage}--\pageref{lastpage}}
\maketitle

\begin{abstract}
The recent detection of very high energy (VHE) emissions from flat spectrum radio quasars (FSRQs) at high redshifts has revealed that the universe is more transparent to VHE $\gamma$-rays than it was expected. It has also questioned the plausible VHE emission mechanism responsible for these objects. Particularly for FSRQs, the $\gamma$-ray emission 
is attributed to the external Compton process (EC).
We perform a detailed spectral study of \emph{Fermi}-detected FSRQ 3C 345 using synchrotron, synchrotron self Compton (SSC) and EC emission mechanisms. The simultaneous data available in optical, UV, X-ray, and $\gamma$-ray energy bands is statistically fitted under these emission mechanisms using the $\chi^2$-minimization technique. Three high flux states and one low flux state are chosen for spectral fitting.
The broadband spectral energy distribution (SED) during these flux states is fitted under different target photon temperatures, and the model VHE 
flux is compared with the 50\hspace{0.05cm}hr CTA sensitivity. Our results indicate a significant VHE emission could be attained during the high flux state from MJD 59635-59715 when the target photon temperature is within 900K to 1200K. Furthermore, our study shows a clear trend of variation in the bulk Lorentz factor of the emission
region as the source transits through different flux states. We also note that during high $\gamma$-ray flux states, an increase in 
external photon temperature demands high bulk Lorentz factors, while this behaviour reverses in case of low $\gamma$-ray flux state.
\end{abstract}




\section{Introduction}
The $\gamma$-ray universe is dominated by blazars \citep{Dermer-Giebels-2016}, a unique class of active galactic nuclei (AGNs) 
having a powerful relativistic jet aligned close to the observer's line of sight \citep{Urry-Padovani-1995}. The broadband 
emission from these sources is highly variable and predominantly non-thermal in nature, and significantly modified due 
to the relativistic Doppler effect \citep{Sambruna-1996, Ulrich-1997, Boettcher-2003, Aharonian-2007, Saito-2013, Raiteri-2013}.
Blazars are further sub-divided into flat spectrum radio quasars (FSRQs), which have broad emission line components in their optical spectrum, and BL Lacaerte objects (BL Lacs) that lack such features \citep{2007ApJ...662..182P}. Besides, FSRQs are more luminous than BL Lacs and hence dominate the high redshift blazar population \citep{Paliya-2017, Ajello-2020}.
The Spectral Energy Distribution (SED) of blazars exhibit two broad peaks \citep{abdo, Giommi} with the low-energy component well 
described by synchrotron emission from a relativistic electron distribution \citep{Blandford-1978, Maraschi-1992, Ghisellini-1993, Hovatta-2009}. Based on the location of synchrotron peak, blazars are further sub-divided into: low-synchrotron-peaked (LSP) 
blazars with peak frequency ($\nu_{\rm syn,peak}< 10^{14}$Hz); intermediate-synchrotron-peaked (ISP) blazars ($10^{14}<\nu_{\rm syn,peak}$ $< 10^{15}$Hz); and high-synchrotron-peaked (HSP) blazars ($\nu_{\rm syn,peak}$ $> 10^{15}$Hz) \citep{Padovani-Giommi-1995, abdo}. The origin of the high energy spectral component is still under debate with models promoting the leptonic origin of the emission
(through inverse Compton processes) \citep{Sahayanathan2012, Shah2019, Shah2021} and others suggest the emission is due to hadronic interactions \citep{Muke-2003, Boettcher-2013}.
Under the leptonic interpretation of the high energy component, the viable target photons for the inverse Compton process are the 
synchrotron photons itself (synchrotron self Compton:SSC) \citep{Konigl-1981, Marscher-1985, Ghisellini-1989} and/or the photons external to the jet (e.g. the thermal photon field from accertion disk/dust, broad line emission, etc.) \citep{1993ApJ...416..458D, 1994ApJ...421..153S, 1996MNRAS.280...67G, 1997A&A...324..395B,  2000ApJ...545..107B, Ghisellini-2009}. The later process is commonly referred as external Compton (EC). In case of HBLs, consideration of SSC process alone is capable to reproduce the high energy spectral component; however, for FSRQs this component is more luminous than the low energy synchrotron component (Compton-dominance) and hence the modelling demands EC process also in addition to SSC. The hadronic interpretation of high energy spectral component involves hadron-initiated cascades, photo-meson/Bethe-Heitler process etc \citep[see,][and references therein for review]{Boettcher-2007}. Besides these, recent models involve the emission from both leptonic and hadronic origin to interpret the high energy emission (lepto-hadronic models) \citep{Diltz-2016, Paliya2016}.

Detection of blazars at high redshift in very high energy (VHE) can provide inputs to understand cosmology. The VHE emission from
distant blazars will undergo significant attenuation due to pair production process with the extragalactic background light (EBL) \citep{Kneiske-2004, Stecker-2006, Franceschini-2008}.
Direct measurement of EBL is cumbersome due to the presence of zodiacal and galactic light, and indirect estimates involving galaxy
counts and cosmological initial conditions are put forth \citep{Kelsall-1998, wright-1998}. However, these estimates depend on the assumptions of the cosmological conditions and can vary when these assumptions are relaxed \citep{Hauser-1998, helgason-2012}. On the other hand, VHE observation of distant blazars can provide strong constraints on these EBL models, provided the source spectrum can be predicted. This prediction has a lot of pitfalls since we have not arrived to any consensus regarding the high energy emission process in blazars. Additionally, the high redshift blazar population is dominated by FSRQs and its interpretation of the high energy spectrum involves multiple emission components, thereby increasing the number of free parameters \citep{Ajello-2020, Hauser-2001, Dwek2013}. Further, only 9 FSRQs are detected in VHE, and this small sample size do not let us to draw a global picture regarding the VHE emission mechanism. 
Simple linear extrapolation of the low energy $\gamma$-ray spectrum of FSRQs to VHE suggests a significant number of sources ($\sim 30$) fall above the sensitivity of Cherenkov Telescope Array Observatory (CTAO), anticipating the increase in source number and better understanding \citep{2022MNRAS.515.4505M}. The extrapolation of low energy $\gamma$-ray to VHE can be questioned since it is optimistic to expect that the same spectral behaviour will persist at these energies. The spectral shape at VHE can differ significantly due to the inverse Compton process happening at the Klein-Nishina regime or the underlying electron distribution will deplete at high energy \citep{Tavecchio_1998, Tavecchio-2008}. Hence, the VHE prediction made from a realistic SED modelling of the source involving simultaneous broadband observations and minimal assumptions can be more reliable.

In this work, we perform a detailed broadband spectral study of the FSRQ 3C\,345 at redshift 0.593 \citep{1965ApJ...142.1667L} using synchrotron and inverse Compton emission processes. This source was already predicted to be a VHE candidate based on the linear extrapolation of \emph{Fermi} spectrum \citep{2022MNRAS.515.4505M}. The simultaneous broadband SED in optical/UV--X-ray--$\gamma$-ray is fitted considering synchrotron, SSC, and EC mechanisms. Particularly, we identify the target photon temperature for which the VHE emission will satisfy the CTAO sensitivity criteria. The paper is organized as follows: in \S \ref{observation}, we report the details of the data reduction procedure, and the analyses are described in \S \ref{temporal}. We discuss the SED model and our results in \S \ref{sed_modeling} and summarize in \S \ref{summary}. Throughout this work, we assumed a cosmology with $H_0$= 71 Km s$^{-1}$ Mpc$^{-1}$, $\Omega_m$ = 0.27, and $\Omega_{\Lambda}$ = 0.73.

\section{Observations and Data Analysis}\label{observation}
In this work, we have used the observations of \emph{Fermi} and \emph{Swift}-XRT/UVOT telescopes. FSRQ 3C\;345 is observed in $\gamma$-ray by \emph{Fermi} since August $2008$. \emph{Fermi}-LAT being a wide-angle telescope, continuous monitoring of the source
was available; however, \emph{Swift} being a pointing telescope, observations of the source are available only during dedicated
flaring states, often triggered by \emph{Fermi} alerts.  For this work, we use all the available observations of the source by \emph{Fermi} and \emph{Swift} starting from August 2008 (MJD 54682) till July 2022 (MJD 59791).

\subsection{\emph{Fermi}-LAT}

The \emph{Fermi}-LAT is a pair conversion telescope built to cover the energy range from 20 MeV to more than 300 GeV \citep{2009ApJ...697.1071A}. It is an outcome of a global partnership between NASA and DOE in the United States and numerous research institutions in France, Italy, Japan, and Sweden. Every 3 hours, \emph{Fermi} scans the entire sky in its normal scanning mode. In this study, $\gamma$-ray data of 3C\,345  is obtained from \emph{Fermi}-LAT during the time period MJD: 54682-59791. To make this data useful for scientific analysis, we processed it with the \emph{Fermitools}\footnote{\url{https://fermi.gsfc.nasa.gov/ssc/data/analysis/documentation/}}--v2.0.1 software. We followed the standard analysis procedure outlined in the \emph{Fermi}-LAT documentation for the data reduction. Specifically, we extracted the P8R3 events from a 15-degree region of interest (ROI) centered on the source location. We selected events with a high probability of being photons using the SOURCE class events with parameters `evclass=128, evtype=3'. Additionally, the photons arriving from the zenith angle $>$ $90^{\circ}$ are blocked to avoid contamination from Earth limb $\gamma$-rays. For spectral analysis, we considered the photons having energy range between 0.1-300 GeV. Also, the latest version 
of \emph{fermipy}-v1.0.1 \citep{Wood2017} is used in the analysis. We modeled the Galactic diffuse emission component with gll$_-$iem$_-$v07.fits and the isotropic emission component with iso$_-$P8R3$_-$CLEAN$_-$V 3$_-$v1.txt. The post-launch instrument response function used in our work is P8R3$_-$SOURCE$_-$V3.

\subsection{\emph{Swift}-XRT}
The X-ray data used in our study were obtained using the \emph{Swift}-XRT instrument on board the Neil Gehrels \emph{Swift} Observatory \citep{Gehrels}. The \emph{Swift} observatory carried a total of 49 observations of the 3C\,345 during the period MJD 54682-59791. The X-ray light curve is obtained such that each \emph{Swift} observation ID corresponds to a single data point in the X-ray light curve. To process the X-ray data collected in photon-counting mode, we employed the XRTDAS V3.0.0 software package, which is part of the HEASOFT package (version 6.27.2). Following the instructions 
in the \emph{Swift}  analysis thread page, standard XRTPIPELINE (Version: 0.13.5) was used to generate the level 2 cleaned event files. The source  events for the spectral analysis were chosen from a circular region of radius
47-arcsec, while the background spectra is chosen from a circular region of radius 100-arcsec. The XIMAGE is utilized to aggregate exposure maps, and the task xrtmkarf is used to create auxiliary response files. Using the task grppha, source spectra are binned so 
that each bin has at least 20 counts. XSPEC version 12.11.0 was used for spectral analysis \citep{xspec}. We fitted the X-ray spectrum with a power-law model after accounting absorption due to neutral hydrogen (Tbabs). The neutral hydrogen column density $\rm n_H$ value was kept fixed at a constant value  $\rm 8.89 \times 10^{19} cm^{-2}$\citep{2005A&A...440..775K}, while the normalization and spectral index of a power-law model were allowed to vary freely.
\subsection{\emph{Swift}-UVOT}
 The \emph{Swift} observatory also provides optical/UV data through its \emph{Swift}-UVOT instrument \citep{Roming-2005}. The \emph{Swift}-UVOT telescope utilizes the filters viz. v, b, u, w1, m2, and w2 to take the observation in optical and UV band \citep{Poole2008}. Again the UVOT data of 3C\,345  is processed into scientific products using the HEASoft package (version 6.26.1). The {\it{uvotsource}} task, included in HEASoft, was employed for image processing. Multiple images obtained through different filters were combined using the \emph{uvotimsum} tool. To extract source counts, a circle of radius of 5 arcsec is chosen as the source region, while a 10 arcsec radius 
source-free circular region is used for the background region. Following \citet{2011ApJ...737..103S}, Galactic extinction using 
$\rm E(B-V)=0.0113$ and $\rm R_v={A_v/E(B-V)}=3.1$ were applied to correct the observed flux.

\section{Temporal Study}\label{temporal}

\begin{figure*}
\centering
\includegraphics[scale=0.60]{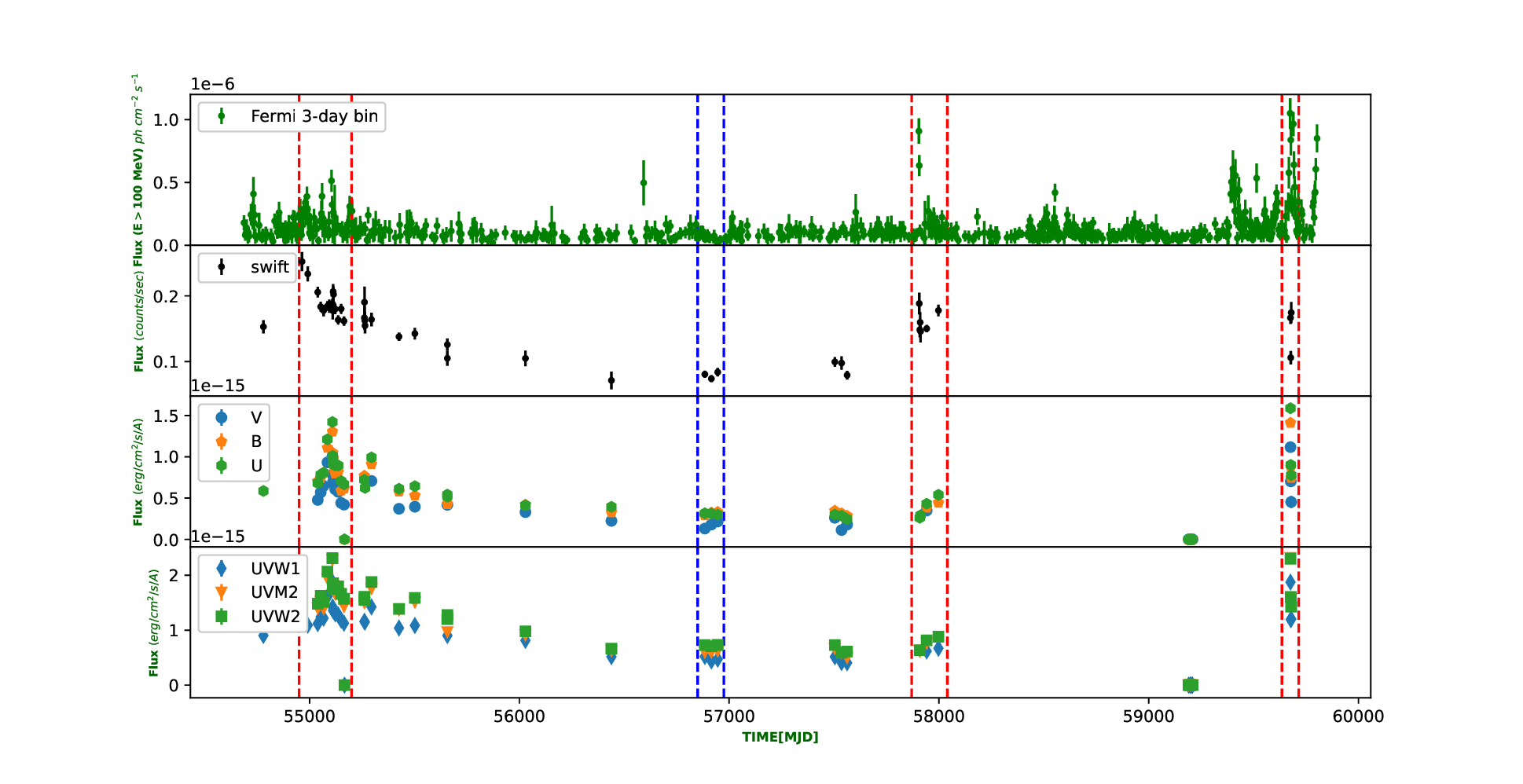}
\caption{Multi-wavelength lightcurve of 3C 345 in different flux states. The top panel of the multiplot displays the 3-day binned $\gamma$-ray lightcurve integrated over the energy range of 0.1–100 GeV, the upper middle panel displays the X-ray lightcurve in the energy range of 0.3-10 keV, the lower middle panel and bottom panel displays UV and Optical lightcurves respectively.}
 \label{fig:Multi_Wavelength lightcurve of 3C345}
\end{figure*}

To examine the temporal behaviour of the source, we obtained the 3-day binned long-term gamma-ray lightcurve of the 
source from the \emph{Fermi}-LAT Light Curve Repository \citep{LCR} during the period  MJD: 54682-59791. The multi-wavelength lightcurve of 3C\,345 obtained using observations from \emph{Fermi}-LAT, \emph{Swift}-XRT/UVOT is shown in Figure~\ref{fig:Multi_Wavelength lightcurve of 3C345}. The $\gamma$-ray lightcurve exhibits many flaring epochs, with simultaneous activity in optical/UV and X-ray energy bands 
available for certain epochs. For three major $\gamma$-ray flaring epochs, simultaneous observations in optical/UV and X-ray bands
are available and these epochs were selected for detailed spectral study (\S \ref{sed_modeling}). Besides these high activity states, simultaneous observations are also available during the quiescent/low activity $\gamma$-ray state of the source. We also choose the low activity epoch during MJD 56850-56975 for the spectral study. These four flux states are demarcated as vertical lines in Figure~\ref{fig:Multi_Wavelength lightcurve of 3C345} and the beginning and the end of these epochs are listed in Table~\ref{tab:start_end_dates}. 

\begin{table*}
\caption{Details of the selected flux states: Col:- 1: States, 2: Starting calendar date, 3: End calendar date, 4: Starting MJD, 5: End MJD, and 6: Nature of the states.}
\Large
     \centering
 \renewcommand{\arraystretch}{2.2}
\begin{tabular}{ l c c c c c}
\hline
\hline
State& Start Date & Stop Date & Start MJD & Stop MJD & Activity \\
\hline 
State I & 2009 April 29 & 2010 Jan 04 & 54950 & 55200 & High Flux \\
State II & 2014 Jul 12 & 2014 Nov 14 & 56850 & 56975 & Low Flux\\
State III & 2017 April 27 & 2017 Oct 14 & 57870 & 58040 &High Flux\\
State IV & 2022 Feb 05 & 2022 May 16 & 59635 & 59715 & High Flux\\
\hline
\hline
\label{tab:start_end_dates}
\end{tabular}
\end{table*}

The maximum $\gamma$-ray activity was observed during MJD 59635-59715 with integrated flux $\rm 1.05\times 10^{-6}$$ \rm ph \hspace{0.1cm}cm^{-2} s^{-1}$ while the X-ray and optical/UV flux during this epoch are $\rm 1.73\times 10^{-1} counts/second$ and $\rm 2.3\times 10^{-15}$ erg $\rm cm^{-2} s^{-1} A^{-1}$. This maximum $\gamma$-ray activity is consistent with the maximum activity in optical/UV; whereas, the X-ray flux from the source reaches a maximum of $\rm 2.5\times 10^{-1} counts/second$ during the epoch MJD 54950-55200. Hence, though correlated flux enhancements can be visualized from the multi-wavelength lightcurve, dissimilar enhancements may indicate different emission processes active at these energy bands. To obtain a better insight into this flux behaviour, we performed a Spearman rank correlation study on the simultaneous observations of the source in the three energy bands. The study resulted in a positive correlation between the $\gamma$-ray and X-ray fluxes with correlation coefficient $\rho=0.44$ and the null hypothesis probability $P<0.005$. Also, a positive correlation is observed between the $\gamma$-ray and optical bands, X-ray and optical bands, and the details of the correlation result between various energy bands are given in Table~\ref{tab:Flux_correlation}. The positive correlation between different energy bands assures that the emission arises from the same region in the 
blazar jet.

\begin{table*}
\caption{Correlation of X-ray light curve with $\gamma$-ray and optical/UV lightcurves using Spearman rank correlation approach. $\rho$ and $\rm P_{value}$ indicates correlation coefficient and probability of null hypothesis respectively. }
\Large
    \centering
    \renewcommand{\arraystretch}{2.0}
    \begin{tabular}{l|c|c|c|c|c}
    \hline
     \hline
     Lightcurves   & $\rho$ & $\rm P_{value}$  & Lightcurves & $\rho$ & $\rm P_{value}$\\
     \hline
       X-ray vs V  & $0.79$ & $< 0.005$ & $\gamma$-ray vs V  & $0.60$ & $< 0.005$\\
       X-ray vs B  & $0.77$ & $< 0.005$ & $\gamma$-ray vs B  & $0.61$ & $< 0.005$  \\
       X-ray vs U  & $0.76$ & $< 0.005$ & $\gamma$-ray vs U  & $0.56$ & $< 0.005$\\
       X-ray vs W1 & $0.75$ & $< 0.005$ & $\gamma$-ray vs W1 & $0.39$ & $0.024$\\
       X-ray vs M2 & $0.71$ & $< 0.005$ & $\gamma$-ray vs M2 & $0.52$ & $< 0.005$ \\
       X-ray vs W2 & $0.75$ & $< 0.005$ & $\gamma$-ray vs W2 & $0.52$ & $< 0.005$\\
       $\gamma$-ray vs X-ray  & $0.44$ & $<0.005$ \\
       \hline
        \hline
       \label{tab:Flux_correlation}
    \end{tabular}
\end{table*}

\begin{figure}
\includegraphics[width=\columnwidth]{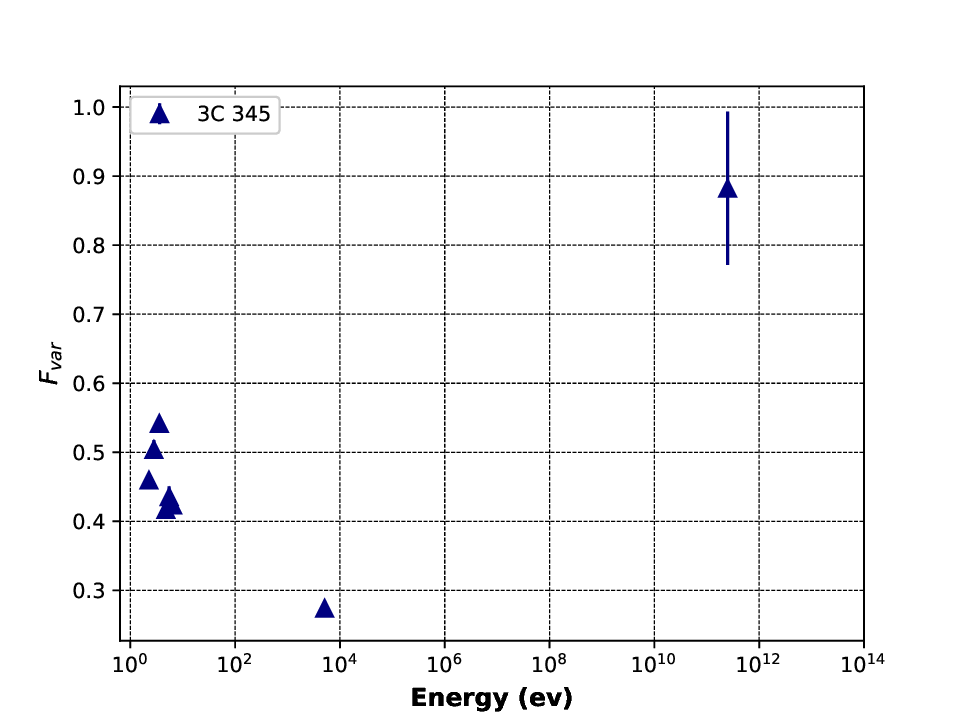}
\caption{Energy dependent Fractional Variability in different energy bands. }
 \label{fig:Fractional variability of 3C 345}
\end{figure}

Though Spearman rank correlation study confirms the correlated flux variability over the optical/UV, X-ray and the $\gamma$-ray bands,
the dissimilar flux enhancements suggest a difference in the variability behaviour of the source. To study this energy-dependent 
variability behaviour, we compute the fractional variability amplitude in different energy bands using \citep{Vaughan}
\begin{equation}
    \label{eq3}
   \rm F_{var}=\sqrt{\frac{S^2-\overline{\sigma_{err}^2}}{\overline{F}^2}}
\end{equation}
Here, $\rm S^2$ denotes the variance, $\rm \overline{F}$ denotes the mean flux and $\rm \overline{\sigma_{err}^2}$ the mean square 
of the measurement error on the flux points. The uncertainty on $\rm F_{var}$ is given by \citep{Vaughan}
\begin{equation}
    \label{eq4}
    \rm F_{var,err}=\sqrt{\frac{1}{2N}\left(\frac{\overline{\sigma_{err}^2}}{F_{var}\overline{F}^2}\right)^2+\frac{1}{N}\frac{\overline{\sigma_{err}^2}}{\overline{F}^2}}
\end{equation}
where, N is the number of simultaneous flux points in the light curve across all energy bands. 
In Figure~\ref{fig:Fractional variability of 3C 345}, we show the plot between $\rm F_{var}$ and the photon energy, and in  
Table~\ref{tab:Fractional_variability} the obtained values of $\rm F_{var}$ with the associated uncertainty for 
UVOT, X-ray, and $\gamma$-ray bands are listed. The lowest values of $\rm F_{var}$ is witnessed for the  X-ray energy
while the highest value corresponds to $\gamma$-ray energy. 

\begin{table}
\caption{Fractional Variability amplitude ($\rm F_{var}$) of source in different energy bands with simultaneous data across the light curve. }
\Large
    \centering
    \renewcommand{\arraystretch}{1.75}
    \begin{tabular}{l|c}
    \hline
     \hline
     Energy band    & $F_{var}$ \\
     \hline
       $\gamma$-ray (0.1–100 GeV)  & $0.88 \pm 0.11$ \\
       X-ray (0.3–10 keV)  & $0.27 \pm 0.01$ \\
       UVW2   & $0.50 \pm 0.01$  \\
       UVM2  & $0.46 \pm 0.01$ \\
       UVW1  & $0.54 \pm 0.01$ \\
       U     & $0.42 \pm 0.01$  \\
       B   &  $0.42 \pm 0.01$\\
       V    & $0.44 \pm 0.01$  \\
       \hline
        \hline
       \label{tab:Fractional_variability}
    \end{tabular}
\end{table}

Such behaviour has also been reported for various other blazars \citep[see, e.g.,][]{Balokovi, Chidiac, Rani, Shah2021, Malik}. 
If we associate the variability behaviour of the source
with changes in the jet parameters, like bulk Lorentz factor, then one would expect similar $\rm F_{var}$ over the
optical/UV and X-ray energy bands (assuming the emission at these energies are due to synchrotron and SSC processes) \citep{Sahayanathan2018}. On the other hand, large variations in $\rm F_{var}$ for the optical/UV and
$\gamma$-ray bands compared to X-ray band can be interpreted in terms of the changes in the emitting electron
distribution. The X-ray emission may fall on the low energy tail of the Compton spectral component, while the 
optical/UV and $\gamma$-ray emission fall on the high energy end of the synchrotron and Compton spectral component (\S \ref{sed_modeling}). Consistently, X-ray emission involves low energy electrons compared to the optical/UV or $\gamma$-ray emission. Since the radiative loss rate scales as the square of the electron energy, high energy electrons lose their energy faster than the low energy ones. This may result in high variability in $\gamma$-ray and optical/UV energies\
compared to X-ray.

\section{Broadband SED Modelling}\label{sed_modeling}
\begin{figure}
	\includegraphics[width=0.33\textwidth,angle=270]{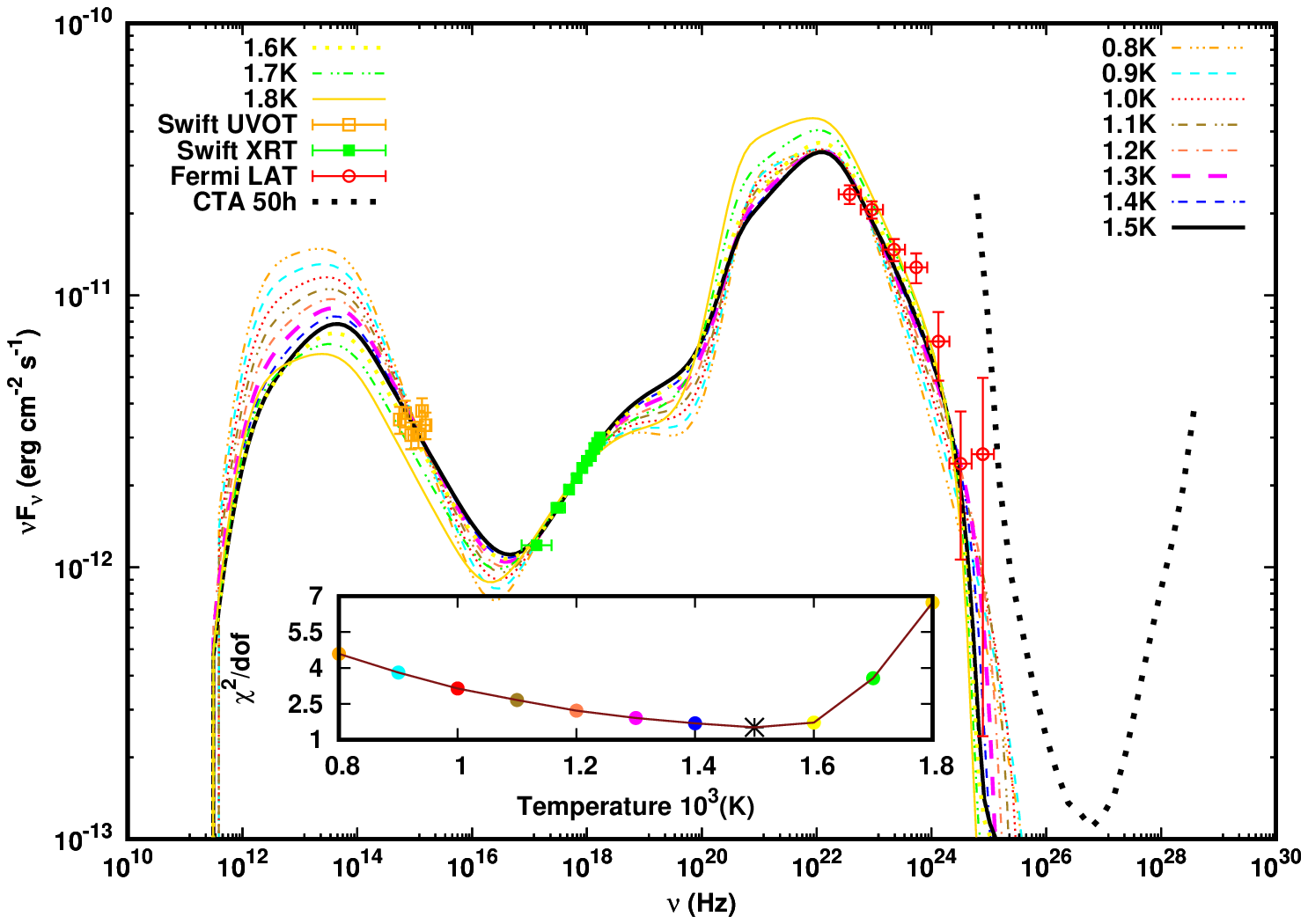}
    \caption{SEDs of 3C 345 during state I (MJD 54950-55200) at various temperatures. The left and right keys illustrate the SEDs at different temperatures. The black dotted curve represents the 50\hspace{0.05cm}hr CTA sensitivity curve. The $\chi^{2}/dof$ versus temperature variation is shown in the inset plot. $*$ corresponds to the temperature at which $\chi^{2}/dof$ is minimum.}
    \label{fig:sed1}
\end{figure}

\begin{figure}
	\includegraphics[width=0.33\textwidth,angle=270]{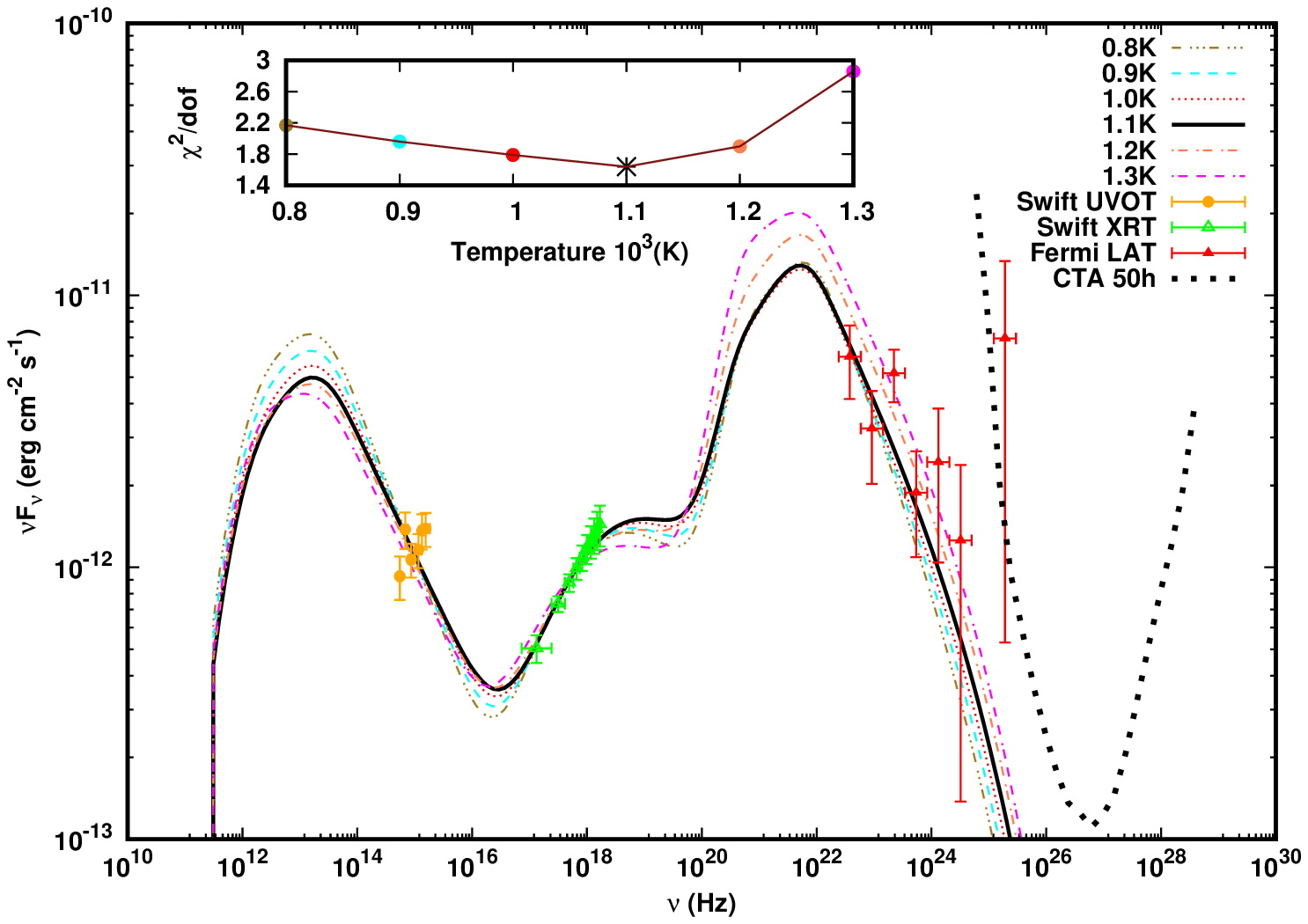}
    \caption{SEDs of 3C 345 during state II (MJD 56850-56975) at various temperatures. The right keys illustrate the SEDs at different temperatures. The black dotted curve represents the 50\hspace{0.05cm}hr CTA sensitivity curve. The $\chi^{2}/dof$ versus temperature variation is shown in the inset plot. $*$ corresponds to the temperature at which $\chi^{2}/dof$ is minimum.}
     \label{fig:sed2}
\end{figure}

\begin{figure}
	\includegraphics[width=0.33\textwidth,angle=270]{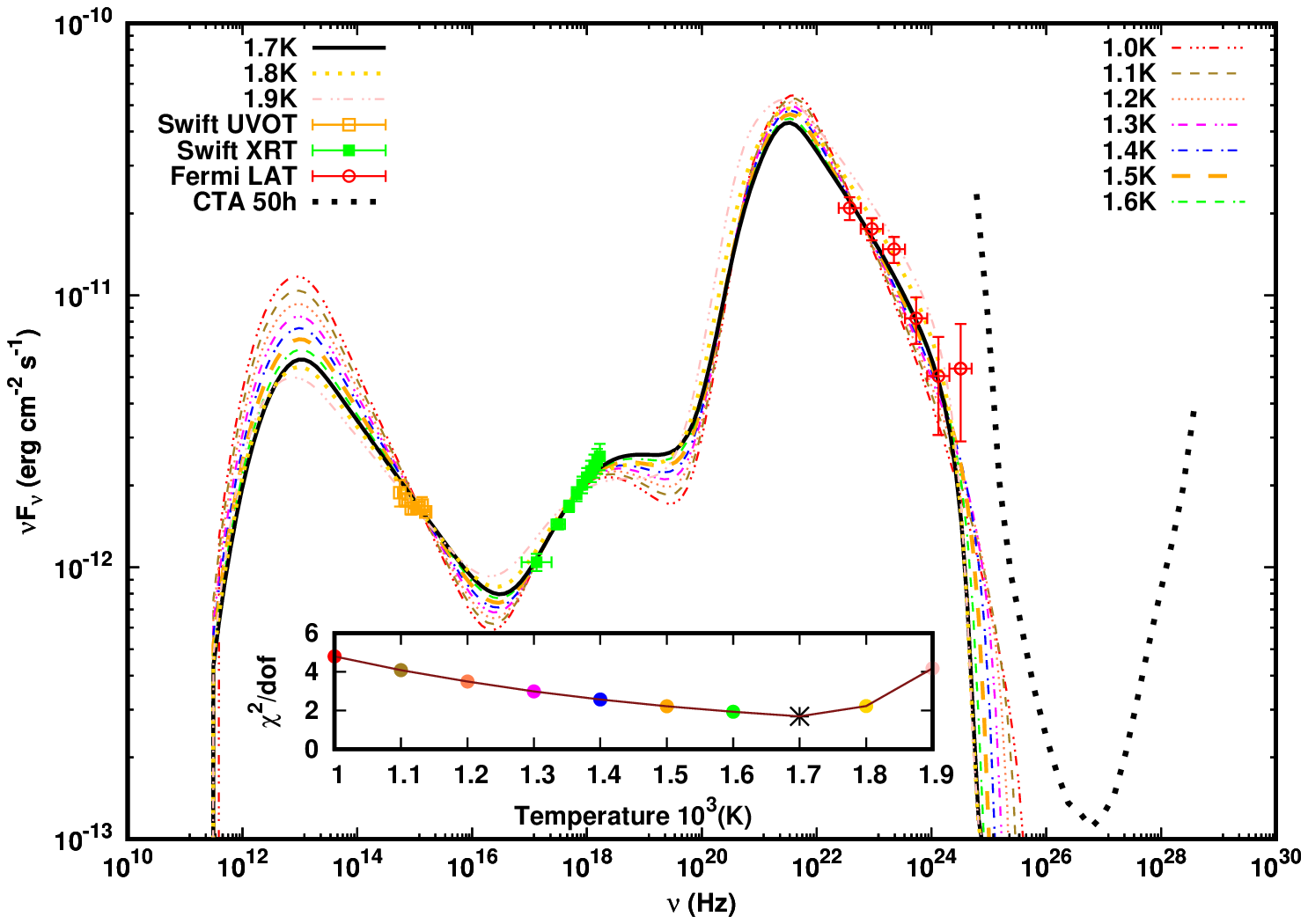}
    \caption{SEDs of 3C 345 during state III (MJD 57870-58040) at various temperatures. The left and right keys illustrate the SEDs at different temperatures. The black dotted curve represents the 50\hspace{0.05cm}hr CTA sensitivity curve. The $\chi^{2}/dof$ versus temperature variation is shown in the inset plot. $*$ corresponds to the temperature at which $\chi^{2}/dof$ is minimum.}
     \label{fig:sed3}
\end{figure}

\begin{figure}
	\includegraphics[width=0.33\textwidth,angle=270]{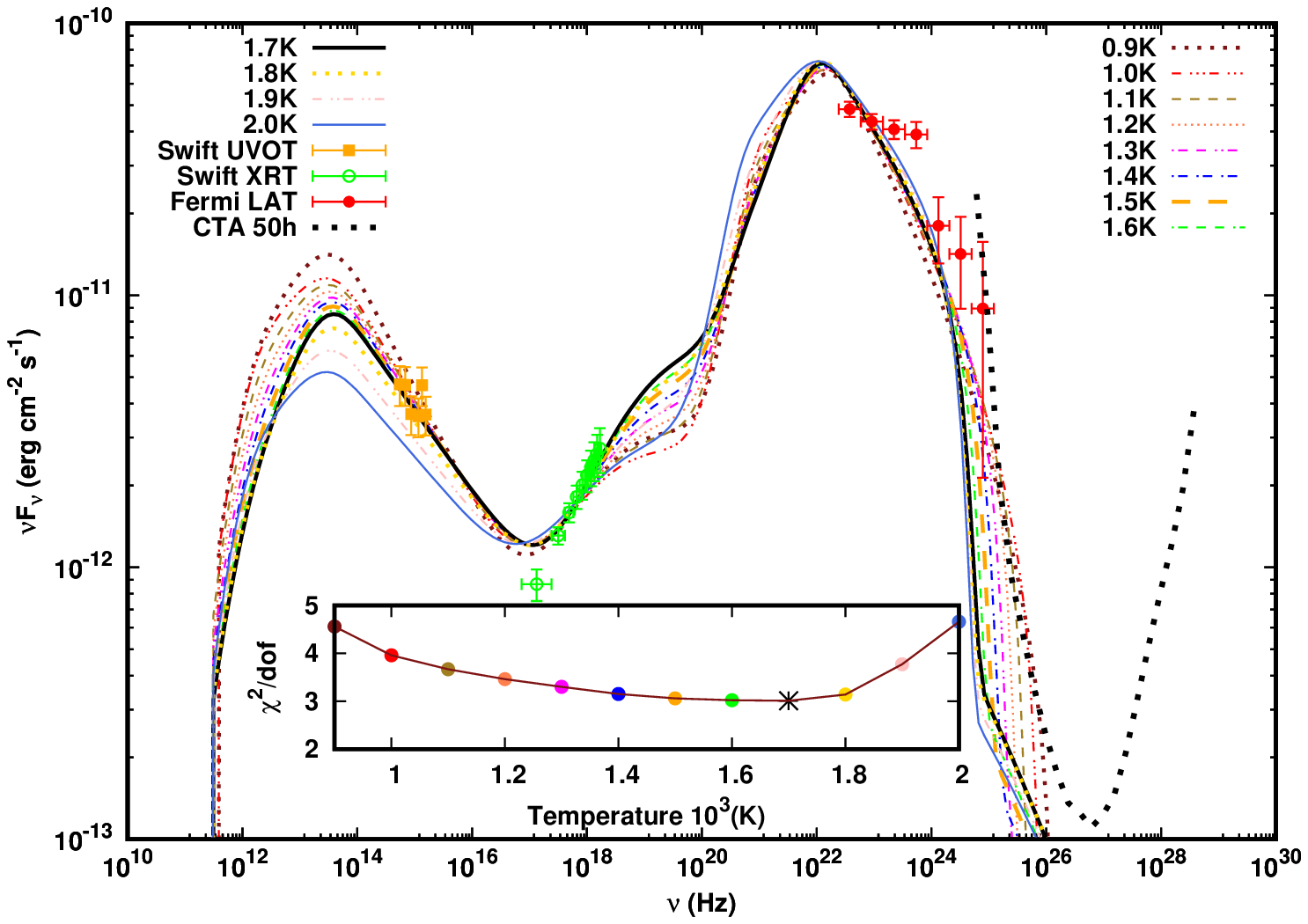}
    \caption{SEDs of 3C 345 during state IV (MJD 59635-59715) at various temperatures. The left and right keys illustrate the SEDs at different temperatures. The black dotted curve represents the 50\hspace{0.05cm}hr CTA sensitivity curve. The $\chi^{2}/dof$ versus temperature variation is shown in the inset plot. $*$ corresponds to the temperature at which $\chi^{2}/dof$ is minimum.}
     \label{fig:sed4}
\end{figure}

\begin{figure}
	\includegraphics[width=0.35\textwidth,angle=270]{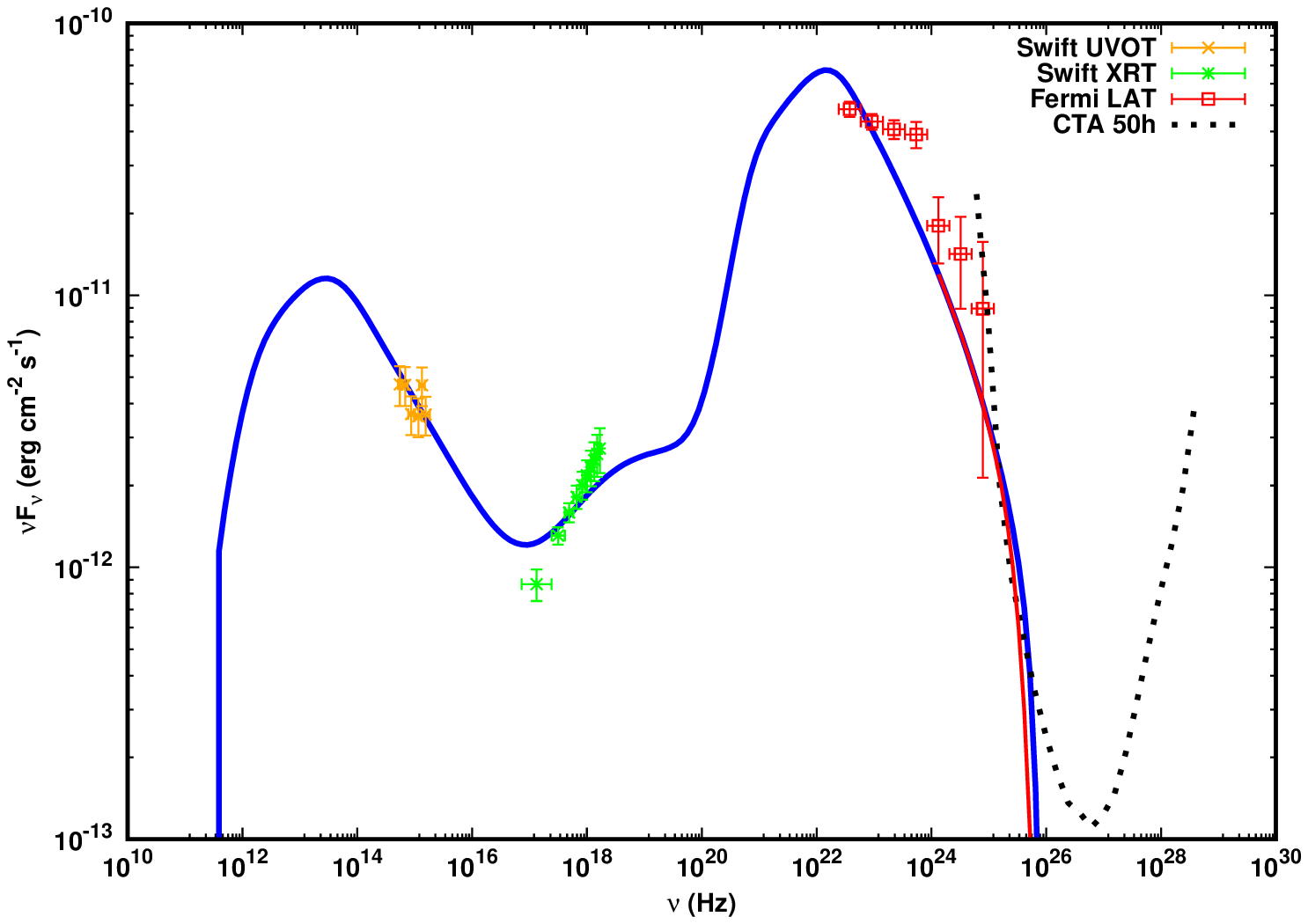}
    \caption{EBL corrected SED of State IV (MJD 59635-59715) at 1000K temperature. The red solid line indicates the EBL-corrected spectrum and the black dotted curve represents the 50\hspace{0.05cm}hr CTA sensitivity curve.}
    \label{fig:EBL}
\end{figure}

To model the broadband SED during different flux states, we performed a spectral fit 
using an one-zone leptonic emission model \citep{Sahayanathan2012, Sahayanathan2018}. Under this model, the emission region is assumed to be a spherical 
region of radius $R$ and populated with a  broken power-law electron distribution given by
\begin{equation}
 N(\gamma)\;d\gamma = K \times
\begin{cases}
\gamma^{-p}d\gamma& {\rm for  } \quad\gamma_{\rm min} <\gamma<\gamma_{\rm b}\\
\gamma^{ q-p}_{\rm b}\gamma^{-q}d\gamma &{\rm for} \quad \gamma_{\rm b} < \gamma < \gamma_{\rm max}
\end{cases} \quad cm^{-3}
\label{equation(3)}
\end{equation}
where $K$ is the normalization factor, $\gamma$ is the dimensionless electron energy with $\gamma_{\rm min}$ and $\gamma_{\rm max}$ 
as the minimum and maximum electron energy, $\gamma_{\rm b}$ the energy at which the power-law distribution breaks, and $p$ and $q$ are the low and high energy particle indices. The emission region is permeated with a tangled magnetic field $B$ and moves down the 
jet with a bulk Lorentz factor $\Gamma$ at an angle $\theta$ with respect to the observer. The electron distribution described
in equation \ref{equation(3)} will lose their energy through synchrotron and inverse Compton radiative processes. The emissivities due to these radiative processes are solved numerically and the flux at the observer is obtained considering the cosmological effects. The relativistic motion of the emission region will also affect the observed flux due to the Doppler factor $\delta = [\Gamma(1-\beta\;{\rm cos}\theta)]^{-1}$. The numerical routine developed is coupled with the  X-ray spectral fitting package XSPEC \citep{xspec}, and the broadband spectral fitting was performed \citep{Sahayanathan2018}. The $\rm n_H$-corrected source X-ray flux is calculated using the tool {\it{cpflux}} in the XSPEC. The optical/UV flux may have additional emission components other than jet emission and often cannot be expressed by a power-law. These flux values will have negligible errors; hence, the broadband spectral fitting will be largely governed by the emission at this energy band. To avoid this bias, we added additional systematic error to the optical data to better represent it by a simple power-law (reduced $\chi^2$ $\sim 1$). For states I, II, and IV, this additional systematic error was chosen to be 11\%, 14\%, and 16\%, whereas the UVOT spectrum of state III can be fitted by the power-law model without incorporating any systematic error. The ASCII data containing the corrected X-ray, optical/UV, and $\gamma$-ray fluxes were then converted into a PHA file using the HEASARC tool {\it{flx2xsp}}.
In order to reduce the number of free parameters, we adapt a minimalistic emission model and found the broadband SED of 3C\,345 can be better reproduced by considering synchrotron, SSC and the EC emission processes. Under these emission processes, the observed spectrum is mainly governed by 10 parameters with $K$, $p$, $q$ and $\gamma_b$ describing the electron distribution, $\Gamma$, $B$, $R$ and $\theta$ describing the macroscopic emission region properties and in addition, two more parameters describing the external photon field frequency and energy density. The external photon field is assumed to be a black body at temperature $T$ and energy density $U_* = f\; U_{bb}$ where $f$ is the fraction of the black body energy density $U_{bb}$ that participates in the inverse Compton scattering process. To impose additional constraints on the parameters, we assume an equipartition between the particle and the magnetic field energy densities. Besides these, we fixed $\gamma_{\rm min}$ and $\gamma_{\rm max}$ at $10^2$ and $10^7$ and assumed the viewing angle to be $2$ degrees. The fraction $f$ is assumed to be 10\%. The minimal information available in optical, X-ray, and $\gamma$-ray bands do not let us to constrain all the parameters, and hence the initial fit is performed with the parameters set free, and for obtaining the confidence intervals only $p$, $q$, $\Gamma$ and $B$ were set as free parameters. Since our aim is to study the VHE property of the source, identifying the most probable $T$ is crucial as the EC process dominates at this energy regime. Hence, we started the fit with the fixed choice of $T$ as 1000 K and obtained the best-fit $\chi$-square. The fit is then repeated for different choices of $T$, and the corresponding minimum $\chi$-squares are obtained. The minimum of all the individual fit minima will correspond to the global minima and give us the most probable $T$ and other source parameters. 
Among all the emission models considered, the one dominant at VHE $\gamma$-rays is the external Compton. Hence, the spectral fitting corresponding to different flux states was performed in two steps. First, we repeated the fitting procedure by varying the target photon temperature and estimated the best fit $\chi$-square for every fit. Second, we compared the VHE spectrum with 50\hspace{0.05cm}hr CTAO sensitivity curve\footnote{\url{https://www.cta-observatory.org/science/cta-performance/}} for different target photon temperatures. For all the flux states, the spectral fit worsens when the temperature was increased beyond 2000 K. Interestingly, this excludes the EC scattering of BLR photons as a plausible mechanism for the $\gamma$-ray emission. At a temperature equivalent to Ly-$\alpha$ emission line (dominant line emission from BLR), the $\gamma$-ray spectrum declines steeply due to Klein-Nishina effects resulting in poor fit statistics. In Figure $\ref{fig:sed1}-\ref{fig:sed4}$, we show the spectral fit for different target photon temperatures corresponding to the selected flux states. The temperature for which minimum $\chi^2$ obtained is provided as inset. The black dotted curve corresponds to 50\hspace{0.05cm}hr CTAO sensitivity. 
To account for the EBL-induced absorption of the source VHE spectrum, we consider the opacity described in \citet{Franceschini-2008}. In Figure $\ref{fig:EBL}$, we show the EBL corrected SED of State IV at 1000K temperature and the comparison with the CTAO sensitivity.
The best-fit parameters corresponding to different target photon temperatures for the selected epochs are given in Table $\ref{tab:Sed_parameters1}-\ref{tab:Sed_parameters4}$. The best-fit parameters let us to estimate the kinetic power of the jet \citep{Sahayanathan2018}.
\begin{equation}
    P_{\rm jet} = \pi R^2\Gamma^2\beta c(U_p + U_e + U_B)
\end{equation}
where $U_p$, $U_e$ and $U_B$ are the proton, electron and the magnetic field energy densities. We assume the protons are cold and
do not participate in the radiative processes (leptonic model); however, their number is assumed to be equal to the number of 
non-thermal electrons (heavy jet). The estimated jet powers from the best-fit parameters are given in column $7$ of Tables $\ref{tab:Sed_parameters1}-\ref{tab:Sed_parameters4}$.
This power is significantly larger than the radiated power (column $8$) and hence, the jet retains most of its energy to launch at kpc/Mpc
scales. If we relax the heavy jet approximation by reducing the number of protons, the jet kinetic power may significantly decrease and the jet will encounter severe deceleration at the blazar zone itself.

During the quiescent state, the best spectral fit was obtained for the target photon temperature $1100$ K. On the other hand, for the 
high flux states corresponding to MJD 54950-5200, MJD 57870-58040 and MJD 59635-59715, the target photon temperatures providing better fit statistics are 1500, 1700, and 1700 K. The best-fit temperature for different flux states is consistent with the dusty environment with dominant silicate and graphite grains \citep{Błażejowski_2000}. This dusty region is heated by the UV radiation from the accretion disk, and hence, the dust content closer to the central black hole will be hotter than the ones away from it \citep{Ghisellini-2009}. If we interpret this in the light of the best-fit temperatures obtained during different flux states, then the location of the emission region during the quiescent state may be farther from the central black hole compared to the high-activity states. Incidentally, the dust temperatures appear to increase with $\gamma$-ray activity when compared with the activity corresponding to the epochs MJD 56850 -56975 and MJD 57870-58040; however, during the maximum $\gamma$-ray activity corresponding to MJD 59635-59715, the temperature is relatively less. This suggests the high $\gamma$-ray activity is associated with the processes happening closer to the central black hole, though the intensity of the activity does not depend on the location of the emission region. This inference should be considered with caution since the best-fit temperatures 
for all the four states are consistent within 1-$\sigma$ tolerance (see inset of Figures $\ref{tab:Sed_parameters1}-\ref{tab:Sed_parameters4})$.

Our study also suggests that the 
source is not bright enough in VHE to be detected by CTAO during low flux/quiescent states. Among the high flux states, for the states
corresponding to MJD 54950-55200 and MJD 57870-58040, the VHE flux is again less compared to CTAO sensitivity for the range of target photon 
temperatures. However, in case of the major $\gamma$-ray flare during MJD 59635-59715, the VHE flux falls above the CTAO sensitivity when the 
target photon temperature ranges from 900-1200 K. Interestingly, the best-fit target photon temperature corresponding to this activity state is less than the other two high activity states. We also find the high activity states demand large bulk Lorentz factors and this may be associated with enhancement of the target photon energy density by $\Gamma^2$. If we assume a situation that the blazar jet expends most
of its bulk energy in the initial stage, then significant deceleration of the jet can be expected with the increasing distance from the 
central black hole. Interestingly, our fit results suggest that the high best-fit temperatures are associated with large jet Lorentz 
factors, consistent with the above inference. These inferences can be further scrutinized with dedicated multi-wavelength observations
of the source with CTAO. Again our study suggest that the FSRQ 3C\,345 may be detected by CTAO during high flux states and this can effectively constrain the blazar jet environment and the jet energetics.

\begin{figure}
	\includegraphics[width=0.33\textwidth,angle=270]{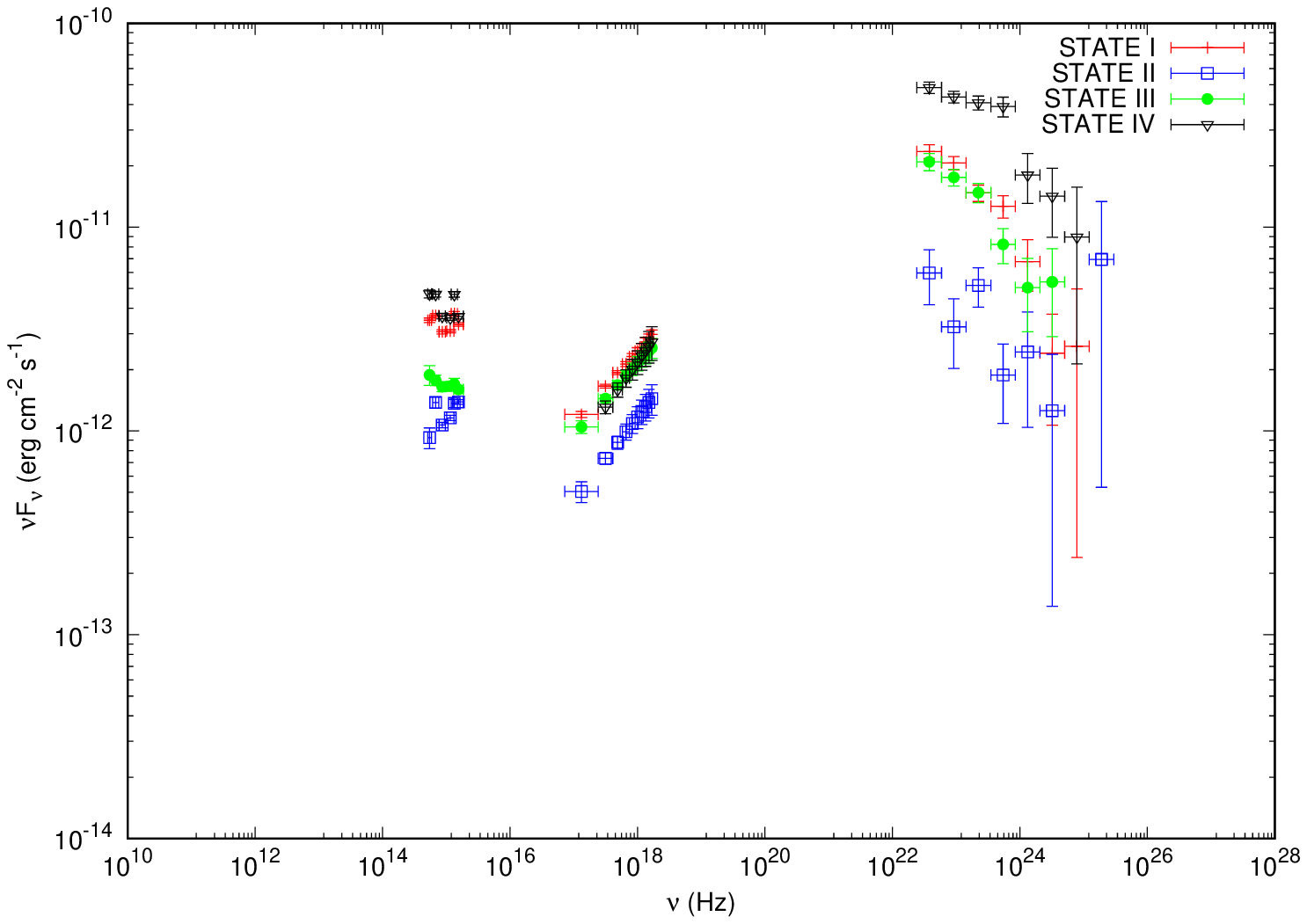}
    \caption{Observed broadband SEDs of 3C\,345 for four time periods, three flaring states are: STATE I (MJD 54950-55200), STATE III (MJD 57870-58040), STATE IV (MJD 59635-59715) and one quiescent STATE II (MJD 56850-56975). The right keys illustrate the SEDs in different states.}
    \label{fig:AllSEDpoints_figure}
\end{figure}
 
\begin{table*}

\caption{Broadband SED model parameters and properties of 3C\,345 for {\bf{STATE I}} are listed in the table below. Col:- 1: Temperature of the target photon field in Kelvin scale, 2: Low energy particle index, 3: High energy particle index, 4: Magnetic field in units of Gauss, 5: Bulk Lorentz factor of the emission region, 6: $\chi^{2}/dof$, 7: Logarithmic jet power in units of erg $S^{-1}$, 8: Logarithmic total radiated power in units of erg $S^{-1}$. The size of the emission region ($R$) is fixed at $7.9\times 10^{15}$cm, Minimum electron Lorentz factor ($\gamma_{\rm min}$) at $10^2$, Maximum electron Lorentz factor ($\gamma_{\rm max}$) at $10^{7}$, Break Lorentz factor ($\gamma_{\rm b}$) at $8.0\times10^2$ and viewing angle ($\theta$) at $2^{\circ}$. The values in subscript and superscript for parameters in the model represent their lower and upper errors, respectively, obtained through the broadband spectral fitting. $--$ symbol indicates that the parameter's upper/lower error value is not constrained.}

\Large
     \centering
    \renewcommand{\arraystretch}{2.5}
		\begin{tabular}{ l c c c c c c c c}
             \hline
              \hline \multirow{2}{*}{} & 
             & \multicolumn{2}{c}{\bf{Free Parameters}} \\ \cline{2-5}
           T& p & q & B & $\Gamma$ & $\chi^{2}/dof$  & $P_{jet}$ & $P_{rad}$ \\ \hline 
			800 & $2.78^{+0.04}_{-0.06}$ & $4.20^{+0.03}_{-0.06}$  & $1.52^{+0.01}_{-0.01}$ & $29.17^{+4.99}_{-4.90}$ 
			& 87.19/19 & 45.72 & 41.85 \\
			900 & $2.73^{+0.07}_{-0.07}$ &  $4.10^{+0.06}_{-0.06}$ & $1.62^{+0.02}_{-0.02}$ & $45.70^{+2.20}_{-2.19}$
			& 72.38/19 & 46.19 & 42.03 \\
           1000 & $2.67^{+0.07}_{-0.07}$ & $4.01^{+0.06}_{-0.06}$ & $1.72^{+0.02}_{-0.02}$ & $55.52^{+2.16}_{-2.08}$ 
			& 59.63/19  & 46.37 & 42.19 \\
          1100 & $2.61^{+0.08}_{-0.08}$ & $3.94^{+0.06}_{-0.05}$ & $1.82^{+0.02}_{-0.02}$ & $64.53^{+2.35}_{-2.19}$ 
			& 50.46/19 & 46.54 & 42.34 \\ 
        1200 & $2.54^{+0.08}_{-0.08}$ & $3.88^{+0.05}_{-0.05}$ & $1.91^{+0.02}_{-0.02}$ & $72.91^{+2.46}_{-2.31}$ 
			& 42.11/19 & 46.68 & 42.49 \\
        
        1300 & $2.48^{+0.09}_{-0.09}$ & $3.82^{+0.05}_{-0.05}$ & $2.01^{+0.03}_{-0.03}$  & $81.11^{+2.71}_{-2.55}$& 36.30/19 & 46.80 & 42.62 \\
        1400 & $2.41^{+0.09}_{-0.09}$ & $3.78^{+0.05}_{-0.04}$  & $2.10^{+0.03}_{-0.03}$  & $89.32^{+2.95}_{-2.78}$ & 32.09/19 & 46.91 & 42.75 \\
        1500 & $2.35^{+0.10}_{-0.10}$ &$3.74^{+0.04}_{-0.04}$  &$2.19^{+0.03}_{-0.03}$  & $97.21^{--}_{--}$ & 28.83/19 & 47.01 & 42.86\\
        1600 & $2.43^{+0.05}_{-0.05}$ &$3.73^{+0.04}_{-0.04}$  &$2.23^{+0.01}_{-0.01}$  & $100^{--}_{--}$ & 33.86/19 & 47.07 & 42.93 \\
        1700 & $2.59^{+0.05}_{-0.05}$ & $3.76^{+0.05}_{-0.04}$  & $2.26^{+0.01}_{-0.01}$  & $100^{--}_{--}$ & 67.92/19 & 47.10 & 42.96 \\
        1800 & $2.75^{+0.05}_{-0.05}$ & $3.78^{+0.05}_{-0.05}$  & $2.28^{+0.01}_{-0.01}$  & $100^{--}_{--}$ & 127.96/19 & 47.13 & 43.00 \\
              \hline
               \hline
              \label{tab:Sed_parameters1}
    	\end{tabular}

  \end{table*}

\begin{table*}

\caption{Broadband SED model parameters and properties of 3C\,345 for {\bf{STATE II}} are listed in the table below. Col:- 1: Temperature of the target photon field in Kelvin scale, 2: Low energy particle index, 3: High energy particle index, 4: Magnetic field in units of Gauss, 5: Bulk Lorentz factor of the emission region, 6: $\chi^{2}/dof$, 7: Logarithmic jet power in units of erg $S^{-1}$, 8: Logarithmic total radiated power in units of erg $S^{-1}$. The size of the emission region ($R$) is fixed at $7.9\times 10^{15}$cm, Minimum electron Lorentz factor ($\gamma_{\rm min}$) at $10^2$, Maximum electron Lorentz factor ($\gamma_{\rm max}$) at $10^{7}$, Break Lorentz factor ($\gamma_{\rm b}$) at $5.0\times10^2$ and viewing angle ($\theta$) at $2^{\circ}$. The values in subscript and superscript for parameters in the model represent their lower and upper errors, respectively, obtained through broadband spectral fitting. $--$ symbol indicates that the parameter's upper/lower error value is not constrained.}

 \Large
     \centering
    \renewcommand{\arraystretch}{2.5}
		\begin{tabular}{ l c c c c c c c c }
   \hline
   \hline \multirow{2}{*}{} & 
             & \multicolumn{3}{c}{\bf{Free Parameters}} \\ \cline{2-5}
           T& p & q & B & $\Gamma$ & $\chi^{2}/dof$ & $P_{jet}$ & $P_{rad}$  \\ \hline
             800 & $2.30^{+0.31}_{-0.33}$ &$4.13^{+0.12}_{-0.12}$  &$1.45^{+0.08}_{-0.05}$  & $15.39^{+2.06}_{-2.05}$ & 41.30/19 & 45.09 & 41.72 \\
			900 & $2.24^{+0.31}_{-0.33}$ & $4.04^{+0.12}_{-0.12}$ & $1.56^{+0.08}_{-0.06}$ & $12.83^{+1.36}_{-1.50}$
			& 37.32/19 & 44.99 & 41.89 \\
           1000 & $2.17^{+0.31}_{-0.34}$ & $3.95^{+0.12}_{-0.12}$ & $1.66^{+0.09}_{-0.06}$ & $11.08^{+1.06}_{-11.08}$ 
			& 33.95/19 & 44.90 & 42.06 \\
          1100 & $2.13^{+0.30}_{-0.20}$ & $3.89^{+0.11}_{-0.09}$ & $1.74^{+0.03}_{-0.05}$ & $10^{+0.69}_{-10.00}$ 
			& 31.25/19 & 44.85 & 42.20 \\ 
        1200 & $2.31^{+0.26}_{-0.28}$ & $3.88^{+0.11}_{-0.11}$ & $1.75^{+0.03}_{-0.03}$ & $10^{+0.20}_{-10.00}$ 
			& 36.04/19 & 44.88 & 42.28 \\
        
        1300 & $2.58^{+0.23}_{-0.24}$ & $3.86^{+0.11}_{-0.11}$ & $1.78^{+0.03}_{-0.03}$  & $10^{+0.10}_{-10.00}$& 54.38/19 & 44.92 & 42.35 \\
         \hline
              \hline
              \label{tab:Sed_parameters2}
		\end{tabular}
  \end{table*}

\vspace{7cm}

\begin{table*}
  \caption{Broadband SED model parameters and properties of 3C\,345 for {\bf{STATE III}} are listed in the table below. Col:- 1: Temperature of the target photon field in Kelvin scale, 2: Low energy particle index, 3: High energy particle index, 4: Magnetic field in units of Gauss, 5: Bulk Lorentz factor of the emission region, 6: $\chi^{2}/dof$, 7: Logarithmic jet power in units of erg $S^{-1}$, 8: Logarithmic total radiated power in units of erg $S^{-1}$. The size of the emission region ($R$) is fixed at $7.9\times 10^{15}$cm, Minimum electron Lorentz factor ($\gamma_{\rm min}$) at $10^2$, Maximum electron Lorentz factor ($\gamma_{\rm max}$) at $10^{7}$, Break Lorentz factor ($\gamma_{\rm b}$) at $3.0\times10^2$ and viewing angle ($\theta$) at $2^{\circ}$. The values in subscript and superscript for parameters in the model represent their lower and upper errors, respectively, obtained through broadband spectral fitting. $--$ symbol indicates that the parameter's upper/lower error value is not constrained.}

   \Large
     \centering
    \renewcommand{\arraystretch}{2.5}
		\begin{tabular}{ l c c c c c c c c}
   \hline
   \hline \multirow{2}{*}{} & 
             & \multicolumn{3}{c}{\bf{Free Parameters}} \\ \cline{2-5}
           T& p & q & B & $\Gamma$ & $\chi^{2}/dof $ & $P_{jet}$ & $P_{rad}$ \\ \hline
             1000 & $1.81^{+0.35}_{-0.41}$ &$3.92^{+0.04}_{-0.04}$  &$1.54^{+0.02}_{-0.02}$  & $45.99^{+2.75}_{-2.67}$ & 86.49/18 & 46.10 & 42.05 \\
             1100 & $1.72^{+0.37}_{-0.40}$ &$3.86^{+0.04}_{-0.04}$  &$1.62^{+0.02}_{-0.02}$  & $54.96^{+2.44}_{-2.52}$ & 73.69/18 & 46.30 & 42.19 \\
			1200 & $1.63^{+0.39}_{-0.41}$ & $3.81^{+0.04}_{-0.04}$ & $1.71^{+0.03}_{-0.02}$ & $62.97^{+2.55}_{-2.44}$
			& 62.94/18 & 46.45 & 42.33 \\
           1300 & $1.55^{+0.40}_{-0.42}$ & $3.76^{+0.04}_{-0.04}$ & $1.79^{+0.03}_{-0.03}$ & $70.56^{+2.65}_{-2.51}$ 
			& 53.87/18 & 46.58 & 42.45 \\
          1400 & $1.46^{+0.41}_{-1.46}$ & $3.71^{+0.04}_{-0.04}$ & $1.87^{+0.03}_{-0.03}$ & $77.94^{+2.64}_{-2.81}$ 
			& 46.27/18 & 46.69 & 42.57 \\ 
        1500 & $1.36^{+0.42}_{-1.36}$ & $3.67^{+0.04}_{-0.04}$ & $1.94^{+0.03}_{-0.03}$ & $85.20^{+2.95}_{-2.76}$ 
			& 39.91/18 & 46.80 & 42.68 \\
        
        1600 & $1.27^{+0.43}_{-1.27}$ & $3.63^{+0.04}_{-0.04}$ & $2.02^{+0.03}_{-0.03}$  & $92.33^{+2.92}_{-2.82}$& 34.71/18 & 46.89 & 42.79 \\
      1700 & $1.15^{+0.45}_{-1.15}$ & $ 3.59^{+0.03}_{-0.04}$  & $ 2.09^{+0.02}_{-0.03}$  & $99.59^{--}_{--}$ & 30.61/18 & 46.98 & 42.88 \\
        1800 & $1.61^{+0.35}_{-0.36}$ & $3.56^{+0.04}_{-0.04}$ & $2.12^{+0.02}_{-0.02}$  &  $100^{--}_{--}$ & 40.08/18 & 47.02 & 42.95\\
        1900 & $2.24^{+0.29}_{-0.30}$ &$3.52^{+0.04}_{-0.04}$  &$2.16^{+0.03}_{-0.02}$  & $100^{--}_{--}$ & 79.36/18 & 47.08 & 42.99 \\
              \hline
               \hline
              \label{tab:Sed_parameters3}
		\end{tabular}
  \end{table*}

\vspace{7cm}
  \begin{table*}
  \caption{Broadband SED model parameters and properties of 3C\,345 for {\bf{STATE IV}} are listed in the table below. Col:- 1: Temperature of the target photon field in Kelvin scale, 2: Low energy particle index, 3: High energy particle index, 4: Magnetic field in units of Gauss, 5: Bulk Lorentz factor of the emission region, 6: $\chi^{2}/dof$, 7: Logarithmic jet power in units of erg $S^{-1}$, 8: Logarithmic total radiated power in units of erg $S^{-1}$. The size of the emission region ($R$) is fixed at $7.9\times 10^{15}$cm, Minimum electron Lorentz factor ($\gamma_{\rm min}$) at $10^2$, Maximum electron Lorentz factor ($\gamma_{\rm max}$) at $10^{7}$, Break Lorentz factor ($\gamma_{\rm b}$) at $6.0\times10^2$ and viewing angle ($\theta$) at $2^{\circ}$. The values in subscript and superscript for parameters in the model represent their lower and upper errors, respectively, obtained through broadband spectral fitting. $--$ symbol indicates that the parameter's upper/lower error value is not constrained.}
  \Large
     \centering
    \renewcommand{\arraystretch}{2.5}
		\begin{tabular}{ l c c c c c c c c }
   \hline
         \hline \multirow{2}{*}{} & 
             & \multicolumn{3}{c}{\bf{Free Parameters}} \\ \cline{2-5}
            
           T& p & q & B & $\Gamma$ & $\chi^{2}/dof $&  $P_{jet}$ & $P_{rad}$ \\ \hline 
			900 & $2.17^{+0.07}_{-0.11}$ & $3.81^{+0.04}_{-0.09}$ & $1.37^{+0.05}_{-0.03}$ & $28.87^{+4.31}_{-3.84}$
			& 86.67/19 & 45.55 & 42.01 \\
           1000 & $2.34^{+0.25}_{-0.26}$ & $3.72^{+0.09}_{-0.10}$ & $1.42^{+0.05}_{-0.05}$ & $40.24^{+4.09}_{-5.43}$ 
			& 75.19/19 & 45.89 & 42.13 \\
          1100 & $2.17^{+0.26}_{-0.27}$ & $3.70^{+0.09}_{-0.09}$ & $1.51^{+0.05}_{-0.06}$ & $51.09^{+3.34}_{-3.66}$ 
			& 69.78/19 & 46.12 & 42.30 \\ 
        1200 & $2.04^{+0.28}_{-0.28}$ & $3.67^{+0.08}_{-0.08}$ & $1.59^{+0.05}_{-0.06}$ & $59.67^{+3.33}_{-3.56}$ 
			& 65.70/19 & 46.29 & 42.45 \\
        
        1300 & $1.89^{+0.29}_{-0.30}$ & $3.65^{+0.08}_{-0.08}$ & $1.67^{+0.05}_{-0.06}$  & $67.74^{+3.46}_{-3.63}$& 62.72/19 & 46.42 & 42.59 \\
        1400 & $1.75^{+0.31}_{-0.31}$ & $3.64^{+0.08}_{-0.08}$  & $1.76^{+0.06}_{-0.06}$  & $75.39^{+3.48}_{-3.67}$ & 59.88/19 & 46.54 & 42.72\\
         1500 & $1.63^{+0.32}_{-0.32}$ & $3.62^{+0.07}_{-0.07}$  & $1.84^{+0.06}_{-0.07}$  & $82.73^{+3.62}_{-3.91}$ & 58.20/19 & 46.64 & 42.84\\
          1600 & $1.51^{+0.34}_{-0.34}$ & $3.60^{+0.07}_{-0.07}$  & $1.92^{+0.07}_{-0.07}$  & $89.91^{+3.81}_{-4.09}$ & 57.39/19 & 46.73 & 42.06 \\
           1700 & $1.39^{+0.35}_{-1.39}$ & $3.59^{+0.07}_{-0.07}$  & $2.01^{+0.07}_{-0.07}$  & $96.98^{--}_{--}$ & 57.19/19 & 46.82 & 43.06\\
           1800 & $1.58^{+0.19}_{-0.13}$ & $3.57^{+0.07}_{-0.07}$ & $2.04^{+0.05}_{-0.06}$  &  $100^{--}_{--}$ & 59.75/19 & 46.88 & 43.12\\
           1900 & $1.97^{+0.12}_{-0.11}$ & $3.53^{+0.07}_{-0.07}$ & $2.04^{+0.05}_{-0.06}$  &  $100^{--}_{--}$ & 71.63/19 & 46.93 & 43.13\\
           2000 & $2.31^{+0.10}_{-0.10}$ & $3.49^{+0.08}_{-0.08}$ & $2.05^{+0.05}_{-0.06}$  &  $100^{--}_{--}$ & 88.56/19 & 46.98 & 43.15\\
              \hline
               \hline
              \label{tab:Sed_parameters4}
		\end{tabular}
  \end{table*}

\section{Summary}\label{summary}
Detection of blazars at high redshift in VHE have cosmological importance. FSRQs dominate the population of blazars at 
high redshift; hence, identifying the plausible VHE FSRQ candidate is desirable. In this work, we study the VHE
property of the FSRQ 3C\,345 through a detailed spectral study involving simultaneous multi-wavelength observation 
in optical/UV, X-ray and $\gamma$-ray energies. Temporal study of the source using multi-wavelength light curve by
\emph{Fermi} and \emph{Swift} suggests the emission at these energy bands are well correlated. This supports that a
single emission region is responsible for the emission at these energies. Further, variability amplitude analysis
suggests the low energy electrons are responsible for the X-ray emission, probably through the SSC process. We chose four epochs with simultaneous observations from the multi-wavelength lightcurves where the source showed different flux activity. Among these four, three epochs correspond to a high activity state in $\gamma$-ray and/or X-ray, while one epoch is chosen as a low activity/quiescent state. The broadband SED corresponding to these flux states can be
well-fitted by synchrotron, SSC, and EC emission processes.

To understand the VHE property of the source, the SED fitting is repeated for different temperatures of the external 
photon field and the fit statistics are compared. The SED fitting for the quiescent state required a relatively
low temperature of the external photon field as well as the bulk Lorentz factor of the jet. This suggests that during the 
low flux state, the emission region is located far from the central black hole, and the jet encounters significant
deceleration at the blazar zone itself. The model VHE flux for all the states except the epoch corresponding to maximum $\gamma$-ray activity fall below the 50 hour CTAO sensitivity. During the epoch of maximum $\gamma$-ray activity, the VHE model flux falls above the detectable range of CTAO when the target photon temperature of the external photon
field is between 900 to 1200 K. These predictions can be verified through future observation of the source by CTAO or by 
the operational imaging atmospheric Cherenkov telescopes, during high $\gamma$-ray activity. Besides, such observations
also have the potential to validate the emission model proposed in this work and scrutinize the parameter space.

\section*{Acknowledgements}

The authors thank the anonymous referee for valuable and insightful comments. AD thanks Malik Zahoor for useful comments, discussions and helpful support. ZS is supported by the Department of Science
and Technology (DST), Govt. of India, under the INSPIRE Faculty grant
(DST/INSPIRE/04/2020/002319). AD also thanks DST for providing financial support. This research has made use of $\gamma$-ray data from {\it Fermi} Science Support Center (FSSC). The work has also used the {\it Swift} Data from the High Energy Astrophysics Science Archive Research Center (HEASARC), at NASA’s Goddard Space Flight Center. 

\section*{Data Availability}
 The data and the model used in this article will be shared on
reasonable request to the corresponding author, Athar Dar
(email: ather.dar6@gmail.com).

\bibliographystyle{mnras}
\bibliography{example} 

\begin{thebibliography}{}
\makeatletter
\relax
\def\mn@urlcharsother{\let\do\@makeother \do\$\do\&\do\#\do\^\do\_\do\%\do\~}
\def\mn@doi{\begingroup\mn@urlcharsother \@ifnextchar [ {\mn@doi@}
  {\mn@doi@[]}}
\def\mn@doi@[#1]#2{\def\@tempa{#1}\ifx\@tempa\@empty \href
  {http://dx.doi.org/#2} {doi:#2}\else \href {http://dx.doi.org/#2} {#1}\fi
  \endgroup}
\def\mn@eprint#1#2{\mn@eprint@#1:#2::\@nil}
\def\mn@eprint@arXiv#1{\href {http://arxiv.org/abs/#1} {{\tt arXiv:#1}}}
\def\mn@eprint@dblp#1{\href {http://dblp.uni-trier.de/rec/bibtex/#1.xml}
  {dblp:#1}}
\def\mn@eprint@#1:#2:#3:#4\@nil{\def\@tempa {#1}\def\@tempb {#2}\def\@tempc
  {#3}\ifx \@tempc \@empty \let \@tempc \@tempb \let \@tempb \@tempa \fi \ifx
  \@tempb \@empty \def\@tempb {arXiv}\fi \@ifundefined
  {mn@eprint@\@tempb}{\@tempb:\@tempc}{\expandafter \expandafter \csname
  mn@eprint@\@tempb\endcsname \expandafter{\@tempc}}}

\bibitem[\protect\citeauthoryear{{Abdo} et~al.,}{{Abdo} et~al.}{2010}]{abdo}
{Abdo} A.~A.,  et~al., 2010, \mn@doi [\apj] {10.1088/0004-637X/716/1/30}, \href
  {https://ui.adsabs.harvard.edu/abs/2010ApJ...716...30A} {716, 30}

\bibitem[\protect\citeauthoryear{{Abdollahi} et~al.,}{{Abdollahi}
  et~al.}{2023}]{LCR}
{Abdollahi} S.,  et~al., 2023, \mn@doi [\apjs] {10.3847/1538-4365/acbb6a},
  \href {https://ui.adsabs.harvard.edu/abs/2023ApJS..265...31A} {265, 31}

\bibitem[\protect\citeauthoryear{{Aharonian} et~al.,}{{Aharonian}
  et~al.}{2007}]{Aharonian-2007}
{Aharonian} F.,  et~al., 2007, \mn@doi [\apjl] {10.1086/520635}, \href
  {https://ui.adsabs.harvard.edu/abs/2007ApJ...664L..71A} {664, L71}

\bibitem[\protect\citeauthoryear{{Ajello} et~al.,}{{Ajello}
  et~al.}{2020}]{Ajello-2020}
{Ajello} M.,  et~al., 2020, \mn@doi [\apj] {10.3847/1538-4357/ab791e}, \href
  {https://ui.adsabs.harvard.edu/abs/2020ApJ...892..105A} {892, 105}

\bibitem[\protect\citeauthoryear{{Arnaud}}{{Arnaud}}{1996}]{xspec}
{Arnaud} K.~A.,  1996, in {Jacoby} G.~H.,  {Barnes} J.,  eds,  Astronomical
  Society of the Pacific Conference Series Vol. 101, Astronomical Data Analysis
  Software and Systems V. p.~17

\bibitem[\protect\citeauthoryear{{Atwood} et~al.,}{{Atwood}
  et~al.}{2009}]{2009ApJ...697.1071A}
{Atwood} W.~B.,  et~al., 2009, \mn@doi [\apj] {10.1088/0004-637X/697/2/1071},
  \href {https://ui.adsabs.harvard.edu/abs/2009ApJ...697.1071A} {697, 1071}

\bibitem[\protect\citeauthoryear{{Balokovi{\'c}} et~al.,}{{Balokovi{\'c}}
  et~al.}{2016}]{Balokovi}
{Balokovi{\'c}} M.,  et~al., 2016, \mn@doi [\apj]
  {10.3847/0004-637X/819/2/156}, \href
  {https://ui.adsabs.harvard.edu/abs/2016ApJ...819..156B} {819, 156}

\bibitem[\protect\citeauthoryear{{Blandford} \& {Rees}}{{Blandford} \&
  {Rees}}{1978}]{Blandford-1978}
{Blandford} R.~D.,  {Rees} M.~J.,  1978, in {Wolfe} A.~M.,  ed., BL Lac
  Objects. pp 328--341

\bibitem[\protect\citeauthoryear{{B{\l}a{\.z}ejowski}, {Sikora}, {Moderski}  \&
  {Madejski}}{{B{\l}a{\.z}ejowski} et~al.}{2000}]{2000ApJ...545..107B}
{B{\l}a{\.z}ejowski} M.,  {Sikora} M.,  {Moderski} R.,   {Madejski} G.~M.,
  2000, \mn@doi [\apj] {10.1086/317791}, \href
  {https://ui.adsabs.harvard.edu/abs/2000ApJ...545..107B} {545, 107}

\bibitem[\protect\citeauthoryear{{Boettcher}, {Mause}  \&
  {Schlickeiser}}{{Boettcher} et~al.}{1997}]{1997A&A...324..395B}
{Boettcher} M.,  {Mause} H.,   {Schlickeiser} R.,  1997, \mn@doi [\aap]
  {10.48550/arXiv.astro-ph/9604003}, \href
  {https://ui.adsabs.harvard.edu/abs/1997A&A...324..395B} {324, 395}

\bibitem[\protect\citeauthoryear{{B{\"o}ttcher} et~al.,}{{B{\"o}ttcher}
  et~al.}{2003}]{Boettcher-2003}
{B{\"o}ttcher} M.,  et~al., 2003, \mn@doi [\apj] {10.1086/378156}, \href
  {https://ui.adsabs.harvard.edu/abs/2003ApJ...596..847B} {596, 847}

\bibitem[\protect\citeauthoryear{{B{\"o}ttcher} et~al.,}{{B{\"o}ttcher}
  et~al.}{2007}]{Boettcher-2007}
{B{\"o}ttcher} M.,  et~al., 2007, \mn@doi [\apj] {10.1086/522583}, \href
  {https://ui.adsabs.harvard.edu/abs/2007ApJ...670..968B} {670, 968}

\bibitem[\protect\citeauthoryear{{B{\"o}ttcher}, {Reimer}, {Sweeney}  \&
  {Prakash}}{{B{\"o}ttcher} et~al.}{2013}]{Boettcher-2013}
{B{\"o}ttcher} M.,  {Reimer} A.,  {Sweeney} K.,   {Prakash} A.,  2013, \mn@doi
  [\apj] {10.1088/0004-637X/768/1/54}, \href
  {https://ui.adsabs.harvard.edu/abs/2013ApJ...768...54B} {768, 54}

\bibitem[\protect\citeauthoryear{Błażejowski, Sikora, Moderski  \&
  Madejski}{Błażejowski et~al.}{2000}]{Błażejowski_2000}
Błażejowski M.,  Sikora M.,  Moderski R.,   Madejski G.~M.,  2000, \mn@doi
  [The Astrophysical Journal] {10.1086/317791}, 545, 107

\bibitem[\protect\citeauthoryear{{Chidiac} et~al.,}{{Chidiac}
  et~al.}{2016}]{Chidiac}
{Chidiac} C.,  et~al., 2016, \mn@doi [\aap] {10.1051/0004-6361/201628347},
  \href {https://ui.adsabs.harvard.edu/abs/2016A&A...590A..61C} {590, A61}

\bibitem[\protect\citeauthoryear{{Dermer} \& {Giebels}}{{Dermer} \&
  {Giebels}}{2016}]{Dermer-Giebels-2016}
{Dermer} C.~D.,  {Giebels} B.,  2016, \mn@doi [Comptes Rendus Physique]
  {10.1016/j.crhy.2016.04.004}, \href
  {https://ui.adsabs.harvard.edu/abs/2016CRPhy..17..594D} {17, 594}

\bibitem[\protect\citeauthoryear{{Dermer} \& {Schlickeiser}}{{Dermer} \&
  {Schlickeiser}}{1993}]{1993ApJ...416..458D}
{Dermer} C.~D.,  {Schlickeiser} R.,  1993, \mn@doi [\apj] {10.1086/173251},
  \href {https://ui.adsabs.harvard.edu/abs/1993ApJ...416..458D} {416, 458}

\bibitem[\protect\citeauthoryear{{Diltz} \& {B{\"o}ttcher}}{{Diltz} \&
  {B{\"o}ttcher}}{2016}]{Diltz-2016}
{Diltz} C.,  {B{\"o}ttcher} M.,  2016, \mn@doi [\apj]
  {10.3847/0004-637X/826/1/54}, \href
  {https://ui.adsabs.harvard.edu/abs/2016ApJ...826...54D} {826, 54}

\bibitem[\protect\citeauthoryear{{Dwek} \& {Krennrich}}{{Dwek} \&
  {Krennrich}}{2013}]{Dwek2013}
{Dwek} E.,  {Krennrich} F.,  2013, \mn@doi [Astroparticle Physics]
  {10.1016/j.astropartphys.2012.09.003}, \href
  {https://ui.adsabs.harvard.edu/abs/2013APh....43..112D} {43, 112}

\bibitem[\protect\citeauthoryear{{Franceschini}, {Rodighiero}  \&
  {Vaccari}}{{Franceschini} et~al.}{2008}]{Franceschini-2008}
{Franceschini} A.,  {Rodighiero} G.,   {Vaccari} M.,  2008, \mn@doi [\aap]
  {10.1051/0004-6361:200809691}, \href
  {https://ui.adsabs.harvard.edu/abs/2008A&A...487..837F} {487, 837}

\bibitem[\protect\citeauthoryear{{Gehrels} et~al.,}{{Gehrels}
  et~al.}{2004}]{Gehrels}
{Gehrels} N.,  et~al., 2004, \mn@doi [\apj] {10.1086/422091}, \href
  {https://ui.adsabs.harvard.edu/abs/2004ApJ...611.1005G} {611, 1005}

\bibitem[\protect\citeauthoryear{{Ghisellini} \& {Madau}}{{Ghisellini} \&
  {Madau}}{1996}]{1996MNRAS.280...67G}
{Ghisellini} G.,  {Madau} P.,  1996, \mn@doi [\mnras] {10.1093/mnras/280.1.67},
  \href {https://ui.adsabs.harvard.edu/abs/1996MNRAS.280...67G} {280, 67}

\bibitem[\protect\citeauthoryear{{Ghisellini} \& {Maraschi}}{{Ghisellini} \&
  {Maraschi}}{1989}]{Ghisellini-1989}
{Ghisellini} G.,  {Maraschi} L.,  1989, \mn@doi [\apj] {10.1086/167383}, \href
  {https://ui.adsabs.harvard.edu/abs/1989ApJ...340..181G} {340, 181}

\bibitem[\protect\citeauthoryear{{Ghisellini} \& {Tavecchio}}{{Ghisellini} \&
  {Tavecchio}}{2009}]{Ghisellini-2009}
{Ghisellini} G.,  {Tavecchio} F.,  2009, \mn@doi [\mnras]
  {10.1111/j.1365-2966.2009.15007.x}, \href
  {https://ui.adsabs.harvard.edu/abs/2009MNRAS.397..985G} {397, 985}

\bibitem[\protect\citeauthoryear{{Ghisellini}, {Padovani}, {Celotti}  \&
  {Maraschi}}{{Ghisellini} et~al.}{1993}]{Ghisellini-1993}
{Ghisellini} G.,  {Padovani} P.,  {Celotti} A.,   {Maraschi} L.,  1993, \mn@doi
  [\apj] {10.1086/172493}, \href
  {https://ui.adsabs.harvard.edu/abs/1993ApJ...407...65G} {407, 65}

\bibitem[\protect\citeauthoryear{{Giommi} et~al.,}{{Giommi}
  et~al.}{2012}]{Giommi}
{Giommi} P.,  et~al., 2012, \mn@doi [\aap] {10.1051/0004-6361/201117825}, \href
  {https://ui.adsabs.harvard.edu/abs/2012A&A...541A.160G} {541, A160}

\bibitem[\protect\citeauthoryear{{Hauser} \& {Dwek}}{{Hauser} \&
  {Dwek}}{2001}]{Hauser-2001}
{Hauser} M.~G.,  {Dwek} E.,  2001, \mn@doi [\araa]
  {10.1146/annurev.astro.39.1.249}, \href
  {https://ui.adsabs.harvard.edu/abs/2001ARA&A..39..249H} {39, 249}

\bibitem[\protect\citeauthoryear{{Hauser} et~al.,}{{Hauser}
  et~al.}{1998}]{Hauser-1998}
{Hauser} M.~G.,  et~al., 1998, \mn@doi [\apj] {10.1086/306379}, \href
  {https://ui.adsabs.harvard.edu/abs/1998ApJ...508...25H} {508, 25}

\bibitem[\protect\citeauthoryear{Helgason \& Kashlinsky}{Helgason \&
  Kashlinsky}{2012}]{helgason-2012}
Helgason K.,  Kashlinsky A.,  2012, The Astrophysical Journal Letters, 758, L13

\bibitem[\protect\citeauthoryear{{Hovatta}, {Valtaoja}, {Tornikoski}  \&
  {L{\"a}hteenm{\"a}ki}}{{Hovatta} et~al.}{2009}]{Hovatta-2009}
{Hovatta} T.,  {Valtaoja} E.,  {Tornikoski} M.,   {L{\"a}hteenm{\"a}ki} A.,
  2009, \mn@doi [\aap] {10.1051/0004-6361:200811150}, \href
  {https://ui.adsabs.harvard.edu/abs/2009A&A...494..527H} {494, 527}

\bibitem[\protect\citeauthoryear{{Kalberla}, {Burton}, {Hartmann}, {Arnal},
  {Bajaja}, {Morras}  \& {P{\"o}ppel}}{{Kalberla}
  et~al.}{2005}]{2005A&A...440..775K}
{Kalberla} P.~M.~W.,  {Burton} W.~B.,  {Hartmann} D.,  {Arnal} E.~M.,  {Bajaja}
  E.,  {Morras} R.,   {P{\"o}ppel} W.~G.~L.,  2005, \mn@doi [\aap]
  {10.1051/0004-6361:20041864}, \href
  {https://ui.adsabs.harvard.edu/abs/2005A&A...440..775K} {440, 775}

\bibitem[\protect\citeauthoryear{{Kelsall} et~al.,}{{Kelsall}
  et~al.}{1998}]{Kelsall-1998}
{Kelsall} T.,  et~al., 1998, \mn@doi [\apj] {10.1086/306380}, \href
  {https://ui.adsabs.harvard.edu/abs/1998ApJ...508...44K} {508, 44}

\bibitem[\protect\citeauthoryear{{Kneiske}, {Bretz}, {Mannheim}  \&
  {Hartmann}}{{Kneiske} et~al.}{2004}]{Kneiske-2004}
{Kneiske} T.~M.,  {Bretz} T.,  {Mannheim} K.,   {Hartmann} D.~H.,  2004,
  \mn@doi [\aap] {10.1051/0004-6361:20031542}, \href
  {https://ui.adsabs.harvard.edu/abs/2004A&A...413..807K} {413, 807}

\bibitem[\protect\citeauthoryear{{Konigl}}{{Konigl}}{1981}]{Konigl-1981}
{Konigl} A.,  1981, \mn@doi [\apj] {10.1086/158638}, \href
  {https://ui.adsabs.harvard.edu/abs/1981ApJ...243..700K} {243, 700}

\bibitem[\protect\citeauthoryear{{Lynds}, {Stockton}  \& {Livingston}}{{Lynds}
  et~al.}{1965}]{1965ApJ...142.1667L}
{Lynds} C.~R.,  {Stockton} A.~N.,   {Livingston} W.~C.,  1965, \mn@doi [\apj]
  {10.1086/148457}, \href
  {https://ui.adsabs.harvard.edu/abs/1965ApJ...142.1667L} {142, 1667}

\bibitem[\protect\citeauthoryear{{Malik}, {Shah}, {Sahayanathan}, {Iqbal}  \&
  {Manzoor}}{{Malik} et~al.}{2022a}]{Malik}
{Malik} Z.,  {Shah} Z.,  {Sahayanathan} S.,  {Iqbal} N.,   {Manzoor} A.,
  2022a, \mn@doi [\mnras] {10.1093/mnras/stac1616}, \href
  {https://ui.adsabs.harvard.edu/abs/2022MNRAS.514.4259M} {514, 4259}

\bibitem[\protect\citeauthoryear{{Malik}, {Sahayanathan}, {Shah}, {Iqbal}  \&
  {Manzoor}}{{Malik} et~al.}{2022b}]{2022MNRAS.515.4505M}
{Malik} Z.,  {Sahayanathan} S.,  {Shah} Z.,  {Iqbal} N.,   {Manzoor} A.,
  2022b, \mn@doi [\mnras] {10.1093/mnras/stac2085}, \href
  {https://ui.adsabs.harvard.edu/abs/2022MNRAS.515.4505M} {515, 4505}

\bibitem[\protect\citeauthoryear{{Maraschi}, {Ghisellini}  \&
  {Celotti}}{{Maraschi} et~al.}{1992}]{Maraschi-1992}
{Maraschi} L.,  {Ghisellini} G.,   {Celotti} A.,  1992, \mn@doi [\apjl]
  {10.1086/186531}, \href
  {https://ui.adsabs.harvard.edu/abs/1992ApJ...397L...5M} {397, L5}

\bibitem[\protect\citeauthoryear{{Marscher} \& {Gear}}{{Marscher} \&
  {Gear}}{1985}]{Marscher-1985}
{Marscher} A.~P.,  {Gear} W.~K.,  1985, \mn@doi [\apj] {10.1086/163592}, \href
  {https://ui.adsabs.harvard.edu/abs/1985ApJ...298..114M} {298, 114}

\bibitem[\protect\citeauthoryear{{M{\"u}cke}, {Protheroe}, {Engel}, {Rachen}
  \& {Stanev}}{{M{\"u}cke} et~al.}{2003}]{Muke-2003}
{M{\"u}cke} A.,  {Protheroe} R.~J.,  {Engel} R.,  {Rachen} J.~P.,   {Stanev}
  T.,  2003, \mn@doi [Astroparticle Physics] {10.1016/S0927-6505(02)00185-8},
  \href {https://ui.adsabs.harvard.edu/abs/2003APh....18..593M} {18, 593}

\bibitem[\protect\citeauthoryear{{Padovani} \& {Giommi}}{{Padovani} \&
  {Giommi}}{1995}]{Padovani-Giommi-1995}
{Padovani} P.,  {Giommi} P.,  1995, \mn@doi [\apj] {10.1086/175631}, \href
  {https://ui.adsabs.harvard.edu/abs/1995ApJ...444..567P} {444, 567}

\bibitem[\protect\citeauthoryear{{Padovani}, {Giommi}, {Landt}  \&
  {Perlman}}{{Padovani} et~al.}{2007}]{2007ApJ...662..182P}
{Padovani} P.,  {Giommi} P.,  {Landt} H.,   {Perlman} E.~S.,  2007, \mn@doi
  [\apj] {10.1086/516815}, \href
  {https://ui.adsabs.harvard.edu/abs/2007ApJ...662..182P} {662, 182}

\bibitem[\protect\citeauthoryear{{Paliya}, {Diltz}, {B{\"o}ttcher}, {Stalin}
  \& {Buckley}}{{Paliya} et~al.}{2016}]{Paliya2016}
{Paliya} V.~S.,  {Diltz} C.,  {B{\"o}ttcher} M.,  {Stalin} C.~S.,   {Buckley}
  D.,  2016, \mn@doi [\apj] {10.3847/0004-637X/817/1/61}, \href
  {https://ui.adsabs.harvard.edu/abs/2016ApJ...817...61P} {817, 61}

\bibitem[\protect\citeauthoryear{{Paliya}, {Marcotulli}, {Ajello}, {Joshi},
  {Sahayanathan}, {Rao}  \& {Hartmann}}{{Paliya} et~al.}{2017}]{Paliya-2017}
{Paliya} V.~S.,  {Marcotulli} L.,  {Ajello} M.,  {Joshi} M.,  {Sahayanathan}
  S.,  {Rao} A.~R.,   {Hartmann} D.,  2017, \mn@doi [\apj]
  {10.3847/1538-4357/aa98e1}, \href
  {https://ui.adsabs.harvard.edu/abs/2017ApJ...851...33P} {851, 33}

\bibitem[\protect\citeauthoryear{{Poole} et~al.,}{{Poole}
  et~al.}{2008}]{Poole2008}
{Poole} T.~S.,  et~al., 2008, \mn@doi [\mnras]
  {10.1111/j.1365-2966.2007.12563.x}, \href
  {https://ui.adsabs.harvard.edu/abs/2008MNRAS.383..627P} {383, 627}

\bibitem[\protect\citeauthoryear{{Raiteri} et~al.,}{{Raiteri}
  et~al.}{2013}]{Raiteri-2013}
{Raiteri} C.~M.,  et~al., 2013, \mn@doi [\mnras] {10.1093/mnras/stt1672}, \href
  {https://ui.adsabs.harvard.edu/abs/2013MNRAS.436.1530R} {436, 1530}

\bibitem[\protect\citeauthoryear{{Rani}, {Krichbaum}, {Lee}, {Sokolovsky},
  {Kang}, {Byun}, {Mosunova}  \& {Zensus}}{{Rani} et~al.}{2017}]{Rani}
{Rani} B.,  {Krichbaum} T.~P.,  {Lee} S.~S.,  {Sokolovsky} K.,  {Kang} S.,
  {Byun} D.~Y.,  {Mosunova} D.,   {Zensus} J.~A.,  2017, \mn@doi [\mnras]
  {10.1093/mnras/stw2342}, \href
  {https://ui.adsabs.harvard.edu/abs/2017MNRAS.464..418R} {464, 418}

\bibitem[\protect\citeauthoryear{{Roming} et~al.,}{{Roming}
  et~al.}{2005}]{Roming-2005}
{Roming} P. W.~A.,  et~al., 2005, \mn@doi [\ssr] {10.1007/s11214-005-5095-4},
  \href {https://ui.adsabs.harvard.edu/abs/2005SSRv..120...95R} {120, 95}

\bibitem[\protect\citeauthoryear{{Sahayanathan} \& {Godambe}}{{Sahayanathan} \&
  {Godambe}}{2012}]{Sahayanathan2012}
{Sahayanathan} S.,  {Godambe} S.,  2012, \mn@doi [\mnras]
  {10.1111/j.1365-2966.2011.19829.x}, \href
  {https://ui.adsabs.harvard.edu/abs/2012MNRAS.419.1660S} {419, 1660}

\bibitem[\protect\citeauthoryear{{Sahayanathan}, {Sinha}  \&
  {Misra}}{{Sahayanathan} et~al.}{2018}]{Sahayanathan2018}
{Sahayanathan} S.,  {Sinha} A.,   {Misra} R.,  2018, \mn@doi [Research in
  Astronomy and Astrophysics] {10.1088/1674-4527/18/3/35}, \href
  {https://ui.adsabs.harvard.edu/abs/2018RAA....18...35S} {18, 035}

\bibitem[\protect\citeauthoryear{{Saito}, {Stawarz}, {Tanaka}, {Takahashi},
  {Madejski}  \& {D'Ammando}}{{Saito} et~al.}{2013}]{Saito-2013}
{Saito} S.,  {Stawarz} {\L}.,  {Tanaka} Y.~T.,  {Takahashi} T.,  {Madejski} G.,
    {D'Ammando} F.,  2013, \mn@doi [\apjl] {10.1088/2041-8205/766/1/L11}, \href
  {https://ui.adsabs.harvard.edu/abs/2013ApJ...766L..11S} {766, L11}

\bibitem[\protect\citeauthoryear{{Sambruna}, {Maraschi}  \& {Urry}}{{Sambruna}
  et~al.}{1996}]{Sambruna-1996}
{Sambruna} R.~M.,  {Maraschi} L.,   {Urry} C.~M.,  1996, \mn@doi [\apj]
  {10.1086/177260}, \href
  {https://ui.adsabs.harvard.edu/abs/1996ApJ...463..444S} {463, 444}

\bibitem[\protect\citeauthoryear{{Schlafly} \& {Finkbeiner}}{{Schlafly} \&
  {Finkbeiner}}{2011}]{2011ApJ...737..103S}
{Schlafly} E.~F.,  {Finkbeiner} D.~P.,  2011, \mn@doi [\apj]
  {10.1088/0004-637X/737/2/103}, \href
  {https://ui.adsabs.harvard.edu/abs/2011ApJ...737..103S} {737, 103}

\bibitem[\protect\citeauthoryear{{Shah}, {Jithesh}, {Sahayanathan}, {Misra}  \&
  {Iqbal}}{{Shah} et~al.}{2019}]{Shah2019}
{Shah} Z.,  {Jithesh} V.,  {Sahayanathan} S.,  {Misra} R.,   {Iqbal} N.,  2019,
  \mn@doi [\mnras] {10.1093/mnras/stz151}, \href
  {https://ui.adsabs.harvard.edu/abs/2019MNRAS.484.3168S} {484, 3168}

\bibitem[\protect\citeauthoryear{{Shah}, {Jithesh}, {Sahayanathan}  \&
  {Iqbal}}{{Shah} et~al.}{2021}]{Shah2021}
{Shah} Z.,  {Jithesh} V.,  {Sahayanathan} S.,   {Iqbal} N.,  2021, \mn@doi
  [\mnras] {10.1093/mnras/stab834}, \href
  {https://ui.adsabs.harvard.edu/abs/2021MNRAS.504..416S} {504, 416}

\bibitem[\protect\citeauthoryear{{Sikora}, {Begelman}  \& {Rees}}{{Sikora}
  et~al.}{1994}]{1994ApJ...421..153S}
{Sikora} M.,  {Begelman} M.~C.,   {Rees} M.~J.,  1994, \mn@doi [\apj]
  {10.1086/173633}, \href
  {https://ui.adsabs.harvard.edu/abs/1994ApJ...421..153S} {421, 153}

\bibitem[\protect\citeauthoryear{{Stecker}, {Malkan}  \& {Scully}}{{Stecker}
  et~al.}{2006}]{Stecker-2006}
{Stecker} F.~W.,  {Malkan} M.~A.,   {Scully} S.~T.,  2006, \mn@doi [\apj]
  {10.1086/506188}, \href
  {https://ui.adsabs.harvard.edu/abs/2006ApJ...648..774S} {648, 774}

\bibitem[\protect\citeauthoryear{Tavecchio \& Ghisellini}{Tavecchio \&
  Ghisellini}{2008}]{Tavecchio-2008}
Tavecchio F.,  Ghisellini G.,  2008, \mn@doi [Monthly Notices of the Royal
  Astronomical Society] {10.1111/j.1365-2966.2008.13072.x}, 386, 945

\bibitem[\protect\citeauthoryear{{Tavecchio}, {Maraschi}  \&
  {Ghisellini}}{{Tavecchio} et~al.}{1998}]{Tavecchio_1998}
{Tavecchio} F.,  {Maraschi} L.,   {Ghisellini} G.,  1998, \mn@doi [\apj]
  {10.1086/306526}, \href
  {https://ui.adsabs.harvard.edu/abs/1998ApJ...509..608T} {509, 608}

\bibitem[\protect\citeauthoryear{{Ulrich}, {Maraschi}  \& {Urry}}{{Ulrich}
  et~al.}{1997}]{Ulrich-1997}
{Ulrich} M.-H.,  {Maraschi} L.,   {Urry} C.~M.,  1997, \mn@doi [\araa]
  {10.1146/annurev.astro.35.1.445}, \href
  {https://ui.adsabs.harvard.edu/abs/1997ARA&A..35..445U} {35, 445}

\bibitem[\protect\citeauthoryear{{Urry} \& {Padovani}}{{Urry} \&
  {Padovani}}{1995}]{Urry-Padovani-1995}
{Urry} C.~M.,  {Padovani} P.,  1995, \mn@doi [\pasp] {10.1086/133630}, \href
  {https://ui.adsabs.harvard.edu/abs/1995PASP..107..803U} {107, 803}

\bibitem[\protect\citeauthoryear{{Vaughan}, {Edelson}, {Warwick}  \&
  {Uttley}}{{Vaughan} et~al.}{2003}]{Vaughan}
{Vaughan} S.,  {Edelson} R.,  {Warwick} R.~S.,   {Uttley} P.,  2003, \mn@doi
  [\mnras] {10.1046/j.1365-2966.2003.07042.x}, \href
  {https://ui.adsabs.harvard.edu/abs/2003MNRAS.345.1271V} {345, 1271}

\bibitem[\protect\citeauthoryear{{Wood}, {Caputo}, {Charles}, {Di Mauro},
  {Magill}, {Perkins}  \& {Fermi-LAT Collaboration}}{{Wood}
  et~al.}{2017}]{Wood2017}
{Wood} M.,  {Caputo} R.,  {Charles} E.,  {Di Mauro} M.,  {Magill} J.,
  {Perkins} J.~S.,   {Fermi-LAT Collaboration} 2017, in 35th International
  Cosmic Ray Conference (ICRC2017). p.~824 (\mn@eprint {arXiv} {1707.09551}),
  \mn@doi{10.22323/1.301.0824}

\bibitem[\protect\citeauthoryear{Wright}{Wright}{1998}]{wright-1998}
Wright E.~L.,  1998, The Astrophysical Journal, 496, 1

\makeatother
\end{thebibliography}
\bsp
\label{lastpage}
\end{document}